\documentclass[11pt, reqno]{article}
\usepackage{jheppub}
\usepackage{graphicx}
\usepackage{tikz}
\usepackage{hyperref}
\usetikzlibrary{shapes,arrows,positioning}
\usetikzlibrary{arrows.meta, 
	decorations.markings, 
	positioning} 
\usepackage{mathrsfs}
\usepackage{adjustbox}
\usepackage{xcolor}
\usepackage{subcaption}
\usepackage{booktabs} 

\newcommand{\nn}{\nonumber \\}

\def\newline{{\hspace{15pt}}}

\def\Tr{\mathop{\rm Tr}}

\def\eref#1{(\ref{#1})}

\def\co{\,,}
\def\ed{\,.}

\title{\boldmath Energy Transport in Randomly Coupled Quantum Systems: A Perturbative Approach
	
}

\author[a,b,c]{Tingfei Li,}
\author[d]{Runyu Chen}

\affiliation[a]{College of Physics Science and Technology, Hebei University, Baoding, 071002, China}
\affiliation[b]{Hebei Key Laboratory of High-precision Computation and Application of Quantum Field Theory, Baoding, 071002, China}
\affiliation[c]{Hebei Research Center of the Basic Discipline for Computational Physics, Baoding, 071002, China}
\affiliation[d]{Zhejiang Institute of Modern Physics, Zhejiang University, Hangzhou, 310027, P. R. China }

\emailAdd{tfli@zju.edu.cn}
\emailAdd{chenrunyu@zju.edu.cn}
\abstract{ We study energy transport between two quantum systems coupled through a random interaction. The central feature of our approach is to model the coupling as a Gaussian random matrix, which enables a simple and systematic perturbative expansion. In the large-$N$ limit, we derive explicit expressions for the energy transfer rate and heat conductance up to second order in the coupling strength.
	Using spectral methods and diagrammatic expansions, we obtain the leading- and next-to-leading-order contributions to the energy transfer rate. We illustrate our results through explicit calculations for Gaussian, constant, semicircular, and Gamma densities of states.
}

\begin{document}
	\maketitle 
	\section{Introduction}
	\label{sec:intro}
	\paragraph{Background and Motivation} 
	Transport phenomena provide one of the simplest manifestations of nonequilibrium behavior and offer a powerful framework for studying thermodynamics away from equilibrium \cite{RevModPhys.93.035008,Stefanucci-Leeuwen_2025,Bertini_2021}. They are directly connected to experimental measurements \cite{Mazurek_2016,Micadei_2019,Mazurek_2021,Li2022,Giordani_2023} but also reveal essential characteristics of a system, making them a fundamental probe of dynamical properties \cite{Karrasch_2013,De_Luca_2014,Bertini_2021,Brighi_2025}. For example, classical and quantum variants of the symmetric and asymmetric simple exclusion processes \cite{MALLICK201517,Bauer2019,Bernard_2019,Bernard_2022} provide a platform for analyzing particle and coherence fluctuations in nonequilibrium thermodynamics. Recent work \cite{Hruza_2023} has shown that coherence fluctuations in the quantum symmetric simple exclusion process can be described using free probability theory \cite{mingo2017free,speicher2025lecturenotesfreeprobability}, revealing an elegant link between physical phenomena and mathematical structures.
	
	Analogous to particle transport, energy transport is a ubiquitous nonequilibrium phenomenon.  
	When two quantum systems at different temperatures are brought into contact, energy flows from the hotter subsystem to the colder one, providing a direct manifestation of the second law of thermodynamics \cite{Callen1985ThermodynamicsAA}.  
	The typical time evolution of the energy current exhibits a characteristic multistage structure: an initial rapid rise to a peak, a subsequent decay to a plateau corresponding to a non-equilibrium steady state (NESS), and finally a gradual relaxation to zero as the two subsystems equilibrate.
	
	Previous studies have often modeled the system--bath coupling as a product of local subsystem operators \cite{Bernard_2012, Almheiri:2019jqq, loganayagam2025solvablemodelsheattransport}, in which case the leading-order energy transfer rate is determined solely by the autocorrelation functions of these operators.
	In \cite{Almheiri:2019jqq}, the equilibration dynamics of a system abruptly coupled to a large bath is investigated.  
	The central quantity of interest is the energy curve, which describes the time evolution of the system energy following a sudden turn-on of the system--bath interaction at \( t = 0 \).  
	It is demonstrated that the early-time growth of the system energy is subject to a universal Planckian bound, thereby constraining the functional form of the initial energy rise.
	
	The authors constructed and thoroughly analyzed a minimal model of system--bath thermalization, in which both subsystems are described by Sachdev--Ye--Kitaev (SYK) Hamiltonians \cite{Kitaev2015, Polchinski_2016, Maldacena_2016, Jevicki:2016bwu, jevicki2016bilocalholographysykmodel}, with the bath taken to be significantly larger than the system.  
	Within this framework, the full energy evolution---encompassing the initial increase, the subsequent turnover to energy loss, the intermediate-stage energy depletion, and the final thermalization---is computed numerically.  
	The analysis relies on the Kadanoff--Baym formalism \cite{Eberlein_2017, Bhattacharya_2019}, which necessitates numerical integration of the corresponding real-time equations.

	In general, the energy transfer rate depends on the initial temperatures of the two subsystems, their respective energy spectra, and the specific form of the coupling operators.  
	Nevertheless, certain fundamental features of the energy transfer are universal, i.e., independent of the particular choice of operators.  
	This observation motivates taking an ensemble average over a class of random interactions. Such an average both simplifies the analysis and makes explicit how energy transfer depends on temperature and spectral properties.  
	
	A further advantage of the ensemble-averaging approach is that it facilitates the systematic computation of higher-order corrections.  
	Recent work on energy transport in the SYK model has proposed an inequality constraining energy flow \cite{Almheiri:2019jqq}.  
	While this inequality has been verified at leading order, both analytically and numerically, its validity beyond leading order remains an open question.  
	By evaluating higher-order contributions, one can obtain a more accurate description of realistic equilibration processes and provide a more stringent test of the conjectured energy-flow bound.  
	This constitutes the second central motivation of the present work.

	\begin{figure}[ht]
		\begin{center}
			\begin{tikzpicture}[
				box/.style={draw, rectangle, minimum width=2cm, minimum height=3cm, thick},
				env/.style={rectangle, rounded corners, minimum width=8cm, minimum height=5cm, fill=gray!30, inner sep=10pt},
				flow/.style={red, ->, thick, line width=2pt},
				work/.style={blue, ->, thick, line width=2pt},
				label/.style={midway, above, font=\large},
				]
				
				\node[env] (env) at (1.75,0) {};
				\node[above] at (env.north) {Environment};
				
				\node[box, fill=blue!20] (system1) at (-0.5,0) {System 1};
				\node[box, fill=red!20] (system2) at (4,0) {System 2};
				
				\node[below] at (system1.south) {$T_1$ (Low)};
				\node[below] at (system2.south) {$T_2$ (High)};
				
				\draw[flow] (system2.west) -- (system1.east) node[label] {$\dot{Q}$};
				
				\draw[work] (env.north west) -- (system1.north) node[label, above, sloped] {$\dot{W}_1$};
				\draw[work] (env.north east) -- (system2.north) node[label, above, sloped] {$\dot{W}_2$};
				
			\end{tikzpicture}
			\caption{Schematic of the model. The gray background represents the environment, which exchanges work currents $\dot{W}_1$ and $\dot{W}_2$ with the two subsystems. The quantity $\dot{Q}$ denotes the heat current between the subsystems.}
			\label{fig:model}
		\end{center}
	\end{figure}
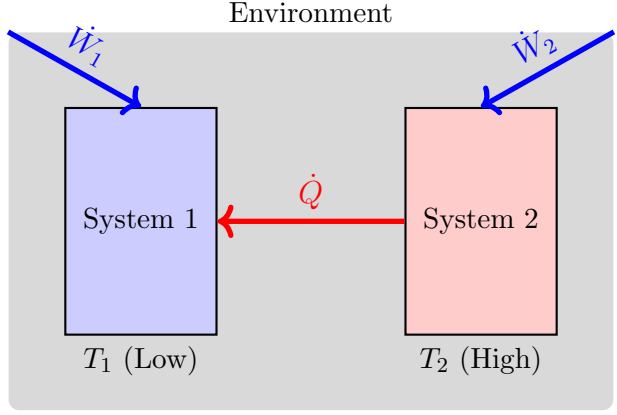
	\paragraph{The model} 
	As shown in Fig.~\ref{fig:model}, we consider two subsystems governed by Hamiltonians $H_1$ and $H_2$, 
	with Hilbert-space dimensions $N_1$ and $N_2$, respectively. Following common treatments in the literature, 
	we may take $N_2$ to be much larger than $N_1$, so that subsystem 2 can be 
	regarded as a large thermal reservoir. For $t>0$, we introduce 
	an interaction $T$ that enables energy exchange between the two systems.
	\begin{align}
		\mathcal{H}=H_1\otimes I^{N_2\times N_2}+I^{N_1\times N_1}\otimes H_2+\theta(t)T
	\end{align}
	where $I^{N\times N}$ denotes the $N$-dimensional identity matrix and $\theta(\cdot)$ is the Heaviside step function. The matrix $T$ has dimensions $N_1 N_2 \times N_1 N_2$. In what follows, we omit explicit identity matrices whenever no confusion can arise. As discussed 
	in the introduction, for calculational simplicity we take $T$ to be a \textit{random 
		matrix} \cite{mehta2004}. Initially, before the interaction is turned on, the two subsystems are prepared in thermal states with inverse temperatures $\beta_1$ and $\beta_2$, respectively. We then examine the energy dynamics of subsystem 1
	\begin{align}
		\mathbb{E}(E_1(t))\equiv \mathbb{E}\left[\text{Tr}\left(H_{1}(t)\rho_{init}\right)\right],~\rho_{init}={1\over Z_1Z_2}e^{-\beta_1 H_1}\otimes e^{-\beta_2 H_2}\ed
	\end{align}
	Here $H_1(t)=U(t)^\dagger H_1 U(t)$, $U(t)=e^{-i \mathcal{H} t}$, and $Z_1,Z_2$ are the initial thermal partition functions of the subsystems
	\begin{align}
		Z_1=\text{Tr}_1(e^{-\beta_1 H_1})\co~Z_2=\text{Tr}_2(e^{-\beta_2 H_2})\ed
	\end{align}
	We denote the partial trace over a subsystem by a subscript on $\Tr$, while the trace without a subscript is taken over the full system. The ensemble average over random matrices is denoted by $\mathbb{E}$.
	In this work, $T$ is taken to be a Gaussian unitary random matrix: 
	\footnote{Time-dependent (Brownian) versions of such matrices are not considered here, as they typically drive the system to infinite temperature and yield trivial dynamics. In contrast, this work focuses on quenched disorder.}
	\begin{align}
		\mathbb{E}(T_{i_1i_2;j_1j_2}T_{k_1k_2;l_1l_2})=J\delta_{i_1l_1}\delta_{i_2l_2}\delta_{j_1k_1}\delta_{j_2k_2}\ed
	\end{align}
	To ensure the model is well-defined, $J$ must scale as $1/(N_1 N_2)$. 
	It is therefore convenient to introduce the renormalized quantities
	\begin{align}
		\mathcal{J}^2 \equiv J \, N_1 N_2\co \qquad 
		\mathcal{Z} \equiv \frac{Z}{N_1 N_2}\ed
	\end{align}
	so that $\mathcal{J}$ has the dimension of energy. Throughout this paper, we work in the diagonal basis of $H_1+H_2$, i.e., the energy eigenbasis. This entails no loss of generality: in any representation, $H$ can be diagonalized by a similarity transformation, under which the interaction becomes $\widetilde{T}=UTU^\dagger$. One can check that $\widetilde{T}$ is again a Gaussian unitary random matrix
	\begin{align}
		\mathbb{E}(\widetilde{T}_{\mu_{1};\mu_{2}}\widetilde{T}_{\nu_{1};\nu_{2}})=J\delta_{\mu_{1}\nu_{2}}\delta_{\mu_{2}\nu_{1}}\ed
	\end{align}
	\paragraph{Main results}
	We focus on the energy current, defined as the time derivative of $E_1$, namely $\mathbb{E}(\dot{E}_1(t))=\dot{W}_1+\dot{Q}$. 
	For convenience, we refer to the second time derivative of $E_1$ as the \textit{energy current acceleration}: $\overline{a}_1 = \mathbb{E}(\ddot{E}_1)$. 
	We consider the thermodynamic limit $N_1, N_2 \to \infty$. In the long-time limit~\footnote{The long-time limit in this paper means first taking the thermodynamic limit, i.e., $N_1, N_2 \to \infty$, and then taking the time $t \to \infty$. This avoids the issue of quantum recurrence.}, the system reaches a non-equilibrium steady state supporting a constant energy (heat) current~\footnote{We will show in the main text that, at least to leading order, the total work done by the environment on the system vanishes in the large-$t$ limit, and the energy flow satisfies the second law of thermodynamics; therefore it can be regarded as a heat current.}. For a sufficiently small initial temperature difference, this current is proportional to the difference. The proportionality constant defines the two-terminal heat conductance $\sigma$ \cite{Gutman_2016}:
	\begin{align}
		\mathbb{E}(\dot{E}_1(\infty)) = \sigma (T_2 - T_1), \qquad 0 < |T_2 - T_1| \ll 1.
	\end{align}
	Our analysis is carried out in the interaction picture, which allows for systematic perturbation theory in $\mathcal{J}$ to arbitrary order. We derive a general expression for $\mathbb{E}(\dot{E}_1(t))$ to second order, which depends only on the \textit{spectrum} and \textit{temperatures} of the two subsystems. In the limit $N_1 = N_2 \to \infty$, we obtain analytical results for several representative spectral densities, including constant and semicircle distributions. We also perform numerical simulations to confirm the validity of the perturbative results.

	\paragraph{Structure of the paper}
	The remainder of this paper is organized as follows.
	Sec.~\ref{sec:heat-work} discusses the distinction between work current and heat current in this model.
	Sec.~\ref{sec:LvNL} presents the perturbative analysis of the energy current. Working in the interaction picture, we develop a diagrammatic expansion in the coupling strength \(\mathcal{J}\). We derive expressions for the energy current and its time derivatives up to second order in \(\mathcal{J}\), both for finite times and in the long-time steady-state limit. 
	In Sec.~\ref{sec:examples}, we provide explicit analytical and numerical results for several representative spectral densities: Gaussian, constant, semicircle, and Gamma distributions. For each case, we compute the energy current and the heat conductance, and in selected examples we compare the perturbative predictions with direct numerical simulations.
	Sec.~\ref{sec:discussion} summarizes our findings, discusses their physical implications including the observed early-time anomalous energy flow, and outlines directions for future research.
	Technical details of the calculations are presented in four appendices.
	
	\paragraph{Remark} 
	A similar model is considered in \cite{chalker2025chaoticmanybodyquantumdynamics}, which studies a chain of $n$ subsystems with random Hamiltonians and random interactions between adjacent sites, with an emphasis on spectral correlations and energy transport. In this sense, our paper corresponds to the $n=2$ case. 
	Refs.~\cite{Gnezdilov_2023,Ohanesjan_2023} also investigated related models, addressing the separation of heat and work and the time evolution of the system's von Neumann entropy. In addition to perturbation theory, one can also integrate out the degrees of freedom of system 2 (the environment) to obtain a dressed master equation for system 1. This enables us to study the system using tools from open quantum systems. As shown in \cite{Tassis_2025}, the steady-state energy current exhibits nontrivial parameter dependence. From the perspective of this work, this behavior can be attributed to the deformation of the system spectrum as parameters vary. 
	
	\section{Separation of Heat and Work}
	\label{sec:heat-work}
	Before proceeding with the perturbative derivation of the energy current, we note from Fig.~\ref{fig:model} that the energy current, i.e., the rate of change of the energy of subsystem 1, contains both the work current from the environment and the heat current between subsystems. We therefore first discuss how to distinguish heat from work. We assume that the Hamiltonian \(H\) is diagonal in the orthonormal basis
	\(\{\lvert i \rangle\}\), i.e.,
	\(H\lvert i \rangle = E_i \lvert i \rangle\).  In this basis any operator \(O\) can be expanded as
	\begin{align}
		O &= \sum_{i,j} O_{ij} \lvert i \rangle\langle j \rvert\ed
	\end{align}
	The commutator with \(H\) then takes the simple form
	\begin{align}
		[H,O] = \sum_{i,j} (E_i - E_j)\, O_{ij}\, \lvert i \rangle\langle j \rvert\ed
		\label{eq:commutator_expansion}
	\end{align}
	From \eqref{eq:commutator_expansion} we see that a matrix element \(O_{ij}\) contributes
	to the commutator if and only if it connects basis states with \emph{different} energies
	(\(E_i \neq E_j\)).  This motivates the following unique decomposition of \(O\):
	\begin{align}
		O_{\parallel} \equiv \sum_{\substack{i,j \\ E_i = E_j}} O_{ij} \lvert i \rangle\langle j \rvert,~
		O_{\perp}   \equiv \sum_{\substack{i,j \\ E_i \neq E_j}} O_{ij} \lvert i \rangle\langle j \rvert\co
	\end{align}
	which clearly satisfy \(O = O_{\parallel} + O_{\perp}\) and
	\begin{align}
		[H,O_{\parallel}] = 0, \qquad
		[H,O_{\perp}] = [H,O]\ed
		\label{eq:commutation_properties}
	\end{align}
	The two parts have the following interpretations:
	\begin{itemize}
		\item \(O_{\parallel}\) retains all matrix elements connecting states of the \emph{same} energy.
		If the spectrum is non-degenerate, \(O_{\parallel}\) is precisely the diagonal part of \(O\);
		with degeneracies it becomes block-diagonal in the energy eigenbasis.
			\item \(O_{\perp}\) contains the matrix elements connecting states of \emph{different} energies
		and carries the entire non-commutativity with \(H\).
	\end{itemize}
We denote the total work current performed by the environment on the system by \(\dot{W}=\dot{W}_1+\dot{W}_2\), and the energy current from subsystem 2 to subsystem 1 by \(\dot{Q}\). Then
\begin{align}\label{eq:dE12}
\dot{E}_1\equiv i\Tr\left(\rho_{init}[\mathcal{H}_0+\mathcal{H}_I, H_1(t)]\right)=\dot{W}_1+ \dot{Q},~\dot{E}_2\equiv i\Tr\left(\rho_{init}[\mathcal{H}_0+\mathcal{H}_I, H_2(t)]\right)=\dot{W}_2- \dot{Q}\ed
\end{align}
The two relations above determine \(\dot{E}_1\) and \(\dot{E}_2\), but an additional relation is needed to isolate \(\dot{Q}\). 
This requires a time-dependent decomposition, even though \(\mathcal{H}=\mathcal{H}_0+\mathcal{H}_I\) itself is time-independent. 
\begin{align}
\mathcal{H}=\mathcal{H}_{\parallel}(t)+\mathcal{H}_{\perp}(t)
\end{align} 
where \([\mathcal{H}_{\parallel}(t),H_1(t)+H_2(t)]=0\) and \(H_i(t)\equiv e^{+i\mathcal{H}t}H_ie^{-i\mathcal{H}t}\). Thus, the work current can be naturally defined as
\begin{align}
\dot{W}_{1}+\dot{W}_{2}&\equiv i\Tr\left(\rho_{init}\left[\mathcal{H},H(t)\right]\right)=i\Tr\left(\rho_{init}\left[\mathcal{H}_{\perp}(t),H_1(t)\right]\right)+i\Tr\left(\rho_{init}\left[\mathcal{H}_{\perp}(t),H_2(t)\right]\right)\ed
\end{align}
Then, from the definition of the energy current in Eq.~\eqref{eq:dE12}, we can read off the heat current
\begin{align}
\dot{Q}=i\Tr\left(\rho_{init}\left[\mathcal{H}_{\parallel}(t),H_{1}(t)\right]\right)=-i\Tr\left(\rho_{init}\left[\mathcal{H}_{\parallel}(t),H_{2}(t)\right]\right)\ed
\end{align}
Although this definition of heat and work currents is natural, its computation is rather involved. At each instant, one must decompose the total Hamiltonian into parts that commute or do not commute with \(H_1(t)+H_2(t)\), which, as discussed above, generally requires diagonalization. Other integral representations can avoid explicit diagonalization, but they introduce different technical complications; for example,
\begin{align}
\mathcal{H}_{\parallel}(t)=\lim_{R\to\infty}{1\over R}\int_{0}^{R}dse^{-iH(t)s}\left(H_{1}+H_{2}+T\right)e^{+iH(t)s}\ed
\end{align}
In short, this rigorous separation of energy transfer and work becomes quite cumbersome in practice.
At the initial time, we have 
\begin{align}
\mathcal{H}_{\perp}(0)=T_{\perp},\qquad \mathcal{H}_{\parallel}(0)=H+T_{\parallel}\co
\end{align}
so that at short times we have the approximation 
\begin{align}
\dot{Q}\approx i\text{Tr}\left(\rho_{init}\left[T_{\parallel},H_{1}(t)\right]\right)\ed
\end{align}

\section{Perturbation Expansion}
\label{sec:LvNL}
To evaluate the time evolution of $H_1$, we work in the interaction picture. For a Hamiltonian decomposed as $\mathcal{H} = \mathcal{H}_0 + \mathcal{H}_I$, the time evolution operator admits the Dyson series expansion
\begin{align}
e^{-i\mathcal{H}t} &= e^{-i\mathcal{H}_{0}t} - i \int_{0}^{t} dt_{1} \, e^{-i\mathcal{H}_{0}(t-t_{1})}\mathcal{H}_{I}e^{-i\mathcal{H}_{0}t_{1}}\nn
&\newline +i^{2} \int_{0}^{t} dt_{1} \int_{0}^{t_{1}} dt_{2} \, e^{-i\mathcal{H}_{0}(t-t_{1})}\mathcal{H}_{I}e^{i\mathcal{H}_{0}(t_{2}-t_{1})}\mathcal{H}_{I}e^{-i\mathcal{H}_{0}t_{2}}+\cdots.
\end{align}
For notational convenience, we introduce the compact expression 
\begin{align}
&(-i)^{n}\int_{0}^{t_{0}=t}dt_{1}\int_{0}^{t_{1}}dt_{2}\ldots\int_{0}^{t_{n-1}}dt_{n}e^{-i\mathcal{H}_{0}(t_{0}-t_{1})}\mathcal{H}_{I}e^{-i\mathcal{H}_{0}(t_{1}-t_{2})}\mathcal{H}_{I}\ldots e^{-i\mathcal{H}_{0}t_{n}}\nn
=&(-i)^{n}\int_{0}^{t}d^nt_>\left(\prod_{k=1}^{n}e^{-i\mathcal{H}_{0}(t_{k-1}-t_{k})}\mathcal{H}_{I}\right)e^{-i\mathcal{H}_{0}t_{n}}\equiv U^{(n)}(t)\ed
\end{align}
where $t_{>}$ denotes time-ordered integration. In our model, we have $\mathcal{H}_0 = H_1 + H_2$ and $\mathcal{H}_I = \theta(t) T$. The expansion for $U^\dagger$ follows analogously.

Since the interaction $T$ is drawn from a Gaussian random matrix ensemble, the ensemble average reduces to pairwise Wick contractions. Consequently, only terms with an even total number of interactions contribute, and we may write

\begin{align}
\mathbb{E}\left[\text{Tr}\left(H_{1}(t)\rho_{init}\right)\right]=\sum_{n=0}^{\infty}\sum_{j+k=2n}\mathbb{E}\left[\text{Tr}\left(U^{\dagger(j)}(t)H_{1}U^{(k)}(t)\rho_{init}\right)\right]
\equiv \sum_{n=0}^\infty  \overline{E}_1^{(n)}(t).
\end{align}
The zeroth-order term reduces to the equilibrium expectation value $E_1^{\beta_1} = \operatorname{Tr}_1(H_1 e^{-\beta_1 H_1})/Z_1$.

\subsection{Leading-order perturbation}
At leading order in perturbation theory, we have 
\begin{align}
\overline{E}_1^{(1)}(t)=\sum_{j+k=2}\mathbb{E}\left[\text{Tr}\left(U^{\dagger(j)}(t)H_{1}U^{(k)}(t)\rho_{init}\right)\right]\ed
\end{align}
To evaluate the ensemble average, we note that for any operator $A$ (an $N_1N_2$-dimensional matrix),  
\begin{align}
\mathbb{E}\left(TAT\right)=J\left(\text{Tr}A\right)\mathbb{I}\ed
\end{align}
Introducing the notation $
\langle O\rangle_{it}\equiv \text{Tr}\left(e^{i\mathcal{H}_{0}t}O\right)
$, we obtain
\begin{align}
\overline{E}_1^{(1)}(t)&=J\int_0^t dt_{1}dt_{2}\langle\rho_{init}\rangle_{it-it_{1}-it_{2}}\langle H_{1}\rangle_{it_{1}+it_{2}-it}\nn
&\newline -J\int_{0}^{t}dt_{1}\int_{0}^{t_{1}}dt_{2}\left[\langle H_{1}\rho_{init}\rangle_{it_{2}-it_{1}}\langle1\rangle_{it_{1}-it_{2}}+\langle H_{1}\rho_{init}\rangle_{it_{1}-it_{2}}\langle1\rangle_{it_{2}-it_{1}}\right]\ed
\end{align}
The traces can be expressed as integrals over the energy spectra of the subsystems. Evaluating the time integrals yields
\begin{align}\label{eq:E1_J1}
\overline{E}_1^{(1)}(t)={\mathcal{J}^2\over \mathcal{Z}}\int[d\epsilon]_{\beta_{1}\beta_{2}}[d\widetilde{\epsilon}]\frac{2-2\cos(t(\widetilde{\epsilon}-\epsilon))}{\left(\widetilde{\epsilon}-\epsilon\right)^{2}}\left(\widetilde{\epsilon}_{1}-\epsilon_{1}\right).
\end{align}
Here we have introduced the shorthand $\epsilon \equiv \epsilon_1 + \epsilon_2$ and $\widetilde{\epsilon} \equiv \widetilde{\epsilon}_1 + \widetilde{\epsilon}_2$. The integration measures are defined as
\begin{align}
[d\epsilon]_{\beta_1\beta_2}=d\epsilon_1d\epsilon_2\rho_1(\epsilon_1)\rho_2(\epsilon_2)e^{-\beta_1 \epsilon_1-\beta_2\epsilon_2}, [d\epsilon]=d\epsilon_1d\epsilon_2\rho_1(\epsilon_1)\rho_2(\epsilon_2)\ed
\end{align}
Differentiating with respect to time gives the leading-order energy current
\begin{align}\label{eq:dE1J1}
\dot{\overline{E}}_{1}^{(1)}(t)=\frac{2\mathcal{J}^2}{\mathcal{Z}}\int[d\epsilon]_{\beta_{1}\beta_{2}}[d\widetilde{\epsilon}]\frac{\sin(t(\widetilde{\epsilon}-\epsilon))}{\widetilde{\epsilon}-\epsilon}\left(\widetilde{\epsilon}_{1}-\epsilon_{1}\right)\ed
\end{align}
In the short-time limit $t \ll 1$, we obtain
\begin{align}
\dot{\overline{E}}_{1}^{(1)}(t\ll 1)={2 \mathcal{J}^2 t\over \mathcal{Z}}\int[d\epsilon]_{\beta_1\beta_2}[d\widetilde{\epsilon}] \left(\widetilde{\epsilon}_{1}-\epsilon_{1}\right)=2\mathcal{J}^2 t\left(E_{1}^{\beta=0}-E_{1}^{\beta=\beta_{1}}\right)> 0\ed 
\end{align}
This indicates an apparently ``anomalous'' early-time behavior: the energy current grows linearly, and consequently the energy of subsystem 1 increases quadratically, even when subsystem 1 is initially hotter than subsystem 2. As discussed in the previous section, the rate of change of a subsystem's energy consists of two contributions: the work current $\dot{W}_1$ from the environment and the heat current $\dot{Q}$ between the subsystems. To determine whether anomalous transport truly occurs, one must therefore examine the heat current rather than the total energy change alone. Moreover, because the environment performs work on the system, the combined dynamics is not that of an isolated bipartite system, and the usual statement of the second law need not apply directly to \(\dot{E}_1\). When the interaction does not commute with the total Hamiltonian, treating $H_1$ and $H_2$ as the proper energy observables of the coupled system may also require care. Thus, even if anomalous heat flow were present, it would not be paradoxical: work performed by the coupling can drive heat from a colder body to a hotter one. In the present random-coupling setting, however, we expect that no anomalous heat flow occurs after ensemble averaging, even when the environment performs work on the system.

This suggests that it is preferable to study interactions that commute with the total Hamiltonian, thereby conserving total energy.  
Such interactions generate time evolution belonging to the class of ``thermal operations,'' which have been extensively investigated in quantum thermodynamics \cite{Brand2013, Lostaglio_2015}. 
Indeed, as shown in \cite{f68k-cjx4, RevModPhys.93.035008}, starting from a thermal initial state and imposing $[T, H_1+H_2]=0$, no anomalous energy flow occurs.  
Anomalous energy flow can only arise in the presence of initial correlations between the two subsystems \cite{PhysRevE.81.061130, PhysRevLett.108.110403, PhysRevE.92.042113,gestsson2026characterizingfunctionalrolequantum}.

Using the decomposition of work current and heat current introduced in the previous section, at early times the heat current is approximately
\begin{align}\label{eq:dotQ_smallt}
\dot{Q}^{(1)}_{t\ll 1}= i\text{Tr}\left(\left[T_{\parallel},H_{1}(t)\right]\right)={2\mathcal{J}^2t\over \mathcal{Z}}\int d[\epsilon]_{\beta_{1}\beta_{2}}d[\eta]\delta(\epsilon-\eta)\left(\eta_{1}-\epsilon_{1}\right)\ed
\end{align}
As shown in Appendix~\ref{appdix:sign}, one can show that $\operatorname{sgn}(\dot{Q}^{(1)}_{t\ll 1})=\operatorname{sgn}(\beta_1-\beta_2)$. Therefore, despite the presence of work current, the early-time heat current defined in this way still obeys the second law of thermodynamics.

We now turn to the late-time behavior. Since the system is taken to have infinitely many degrees of freedom, the non-equilibrium steady state persists indefinitely. This feature is readily apparent in our formulation: in the large-$N$ limit, the spectral density becomes continuous, and in the long-time limit we may apply
\begin{align}
\lim_{t\to \infty} {\sin(t x)\over x}= \pi \delta(x)\co
\end{align}
to obtain the stable energy current
\begin{align}\label{eq:dotE1_larget}
\dot{\overline{E}}^{(1)}_1({\infty})
&=\frac{2\pi\mathcal{J}^2}{\mathcal{Z}}\int d\epsilon_{1}d\epsilon_{2}d\eta\rho_{1}(\epsilon_{1})\rho_{2}(\epsilon_{2})\rho_{1}\left(\epsilon_{1}+\eta\right)\rho_{2}\left(\epsilon_{2}-\eta\right)e^{-\beta_{1}\epsilon_{1}-\beta_{2}\epsilon_{2}}\eta\ed
\end{align}
One can verify that the work current vanishes at leading order in perturbation theory
\begin{align}
\dot{W}^{(1)}(\infty)= \dot{\overline{E}}^{(1)}_1({\infty})+\dot{\overline{E}}^{(1)}_2({\infty})=0\co
\end{align}
so it is natural to regard $\dot{\overline{E}}^{(1)}_1({\infty})$ as the heat current. Moreover, like $\dot{Q}^{(1)}_{t\ll 1}$, the late-time heat current obeys the second law of thermodynamics.  
It is useful to distinguish a finite-size system from the thermodynamic limit. For a finite-size system, the integral in Eq.~\eqref{eq:dE1J1} is replaced by a discrete sum. In the large-$t$ limit, only terms with no net energy change survive. Because we take an ensemble average, quantum recurrences are smoothed out, and the averaged result retains a finite heat current as time tends to infinity. As in the thermodynamic-limit case, one can show that for a discrete spectrum the work current performed by the environment vanishes at infinite time. Nevertheless, some care is required: the perturbative result is valid only for $\mathcal{J}\ll 1$ and $t\ll t_{\text{rec}}\propto \hbar/\Delta$, where $\Delta$ is the typical level spacing. For a large but finite system, the quasi-steady heat-current plateau therefore has a finite lifetime.
\begin{align}
t_{plateau}\propto\frac{N \mathcal{E}}{\mathcal{J}^2}\ll t_{rec}\ed
\end{align}
Here $\mathcal{E}$ is the typical energy scale of the system. In the thermodynamic limit, both $t_{\text{plateau}}$ and $t_{\text{rec}}$ become infinite. It is therefore consistent within the perturbative expansion to take the long-time limit only after the thermodynamic limit; in this order of limits, the final thermal state is not observed. 
To evaluate this integral, we introduce the Fourier representation of the spectral densities:
\begin{align}
\rho(\varepsilon)=\frac{1}{2\pi}\int d\omega e^{+i\omega\varepsilon}\widetilde{\rho}(\omega),\widetilde{\rho}^\beta(\omega)=\int d\varepsilon e^{-i\omega\varepsilon-\beta \varepsilon}\rho(\varepsilon)\co
\end{align}
and denote $\widetilde{\rho}(\omega) = \widetilde{\rho}^{\beta=0}(\omega)$. The steady-state current then becomes
\begin{align}\label{eq:current_infty_w}
\dot{\overline{E}}^{(1)}_{1}(\infty)=\frac{\mathcal{J}^2}{\mathcal{Z}}\frac{i}{2}\int d^{4}\omega\hat{D}_{\omega}\left[\widetilde{\rho}_{1}^{\beta_{1}}(\omega_{1})\widetilde{\rho}_{2}^{\beta_{2}}(\omega_{2})\widetilde{\rho}_{1}\left(\omega_{3}\right)\widetilde{\rho}_{2}\left(\omega_{4}\right)\right]\delta_{\omega_{1}+\omega_{3}}\delta_{\omega_{2}+\omega_{4}}\delta_{\omega_{3}-\omega_{4}}
\end{align}
where $\hat{D}_{\omega}=\left(\partial_{\omega_{3}}-\partial_{\omega_{4}}-\partial_{\omega_{1}}+\partial_{\omega_{2}}\right)$.  
To obtain the heat conductance, we set $\beta_1 = \beta + \delta\beta$, $\beta_2 = \beta$ and expand to first order in $\delta\beta$, yielding
\begin{align}\label{eq:sigma_w}
\sigma=\frac{\mathcal{J}^2}{\mathcal{Z}}\frac{\beta^{2}}{2}\int d^4\omega\partial_{\omega_1}\hat{D}_{\omega}\left[\widetilde{\rho}_{1}^{\beta}(\omega_{1})\widetilde{\rho}_{2}^{\beta}(\omega_{2})\widetilde{\rho}_{1}\left(\omega_{3}\right)\widetilde{\rho}_{2}\left(\omega_{4}\right)\right]\delta_{\omega_{1}+\omega_{3}}\delta_{\omega_{2}+\omega_{4}}\delta_{\omega_{3}-\omega_{4}}+\mathcal{O}(\mathcal{J}^4)\ed
\end{align}
In Appendix~\ref{appdix:asym}, we discuss the asymptotic behavior of the two-point heat conductance in the high- and low-temperature limits; these behaviors can be verified by the explicit examples presented in the main text.

For general time $t$, Eq.~\eref{eq:dE1J1} involves a fourfold integral. 
Because the integration variables appear in the denominator, the Fourier transform cannot be applied directly. We therefore differentiate the energy current once more with respect to time to eliminate the denominator:
\begin{equation}\label{eq:ddot_E1}
\begin{aligned}
	\ddot{\overline{E}}_{1}(t)&=\frac{2\mathcal{J}^2}{\mathcal{Z}}\int[d\epsilon][d\widetilde{\epsilon}]e^{-\beta_{1}\epsilon_{1}-\beta_{2}\epsilon_{2}}\cos\left(t\left(\widetilde{\epsilon}_{1}+\widetilde{\epsilon}_{2}-\epsilon_{1}-\epsilon_{2}\right)\right)\left(\widetilde{\epsilon}_{1}-\epsilon_{1}\right)\\&=\frac{i\mathcal{J}^2}{\mathcal{Z}}\left(\partial_{\omega_{3}}-\partial_{\omega_{1}}\right)\left[\widetilde{\rho}_{1}^{\beta_{1}}(\omega_{1})\widetilde{\rho}_{2}^{\beta_{2}}(\omega_{2})\widetilde{\rho}_{1}\left(\omega_{3}\right)\widetilde{\rho}_{2}\left(\omega_{4}\right)\right]\bigg|_{\omega_{1,2}\to t,\omega_{3,4}\to-t}+\left(t\leftrightarrow-t\right).
\end{aligned}
\end{equation}
The resulting energy current acceleration is expressed entirely in terms of Fourier transforms of the spectral functions. The energy current itself is then recovered by a single time integration: $\dot{\overline{E}}_1(t) = \int_0^t \ddot{\overline{E}}_1(\tau) d\tau$. This approach reduces the original four-fold integral to a single fold, facilitating both theoretical analysis and numerical evaluation.
Finally, we define the integrated energy flux
\begin{align}\label{eq:integrable-flow}
F_\kappa = \int_0^\infty dt \, e^{-\kappa t} \, \dot{E}_1(t)=\frac{1}{\kappa}\int_{0}^{\infty}d\tau e^{-\kappa\tau}\ddot{E}_{1}(\tau)\ed
\end{align}
As established in \cite{Almheiri:2019jqq}, to leading order in perturbation theory the integrated energy flux \(F_{\kappa}\) is guaranteed to be positive whenever the interaction takes the product form \(T = O_1 O_2\), where \(O_1\) and \(O_2\) act on the two subsystems, respectively. Specifically, one has
\begin{align}
\kappa \geq 2 / \beta_1 \quad \Longrightarrow \quad F_\kappa \geq 0,
\end{align}
and we may assume without loss of generality that \(\beta_1 < \beta_2\).

In the present paper, however, the interaction is taken to be a Gaussian random matrix, which cannot be factorized into a simple product of subsystem operators. Consequently, the method used in \cite{Almheiri:2019jqq}, namely constructing a manifestly positive integrand, does not directly apply, and a proof of the same inequality is not available for our model. We nevertheless expect that the inequality continues to hold, but we do not pursue this point further, as \(F_\kappa\) is not the primary focus of our study.
\begin{figure}[ht]
\begin{center}
	\begin{tikzpicture}[
		node distance=2cm,
		tensor/.style={circle, draw=black, thick, fill=white, inner sep=2pt, minimum size=8mm},
		op/.style={rectangle, draw=black, thick, fill=white, inner sep=2pt, minimum size=8mm},
		arcarrow/.style={
			thick,
			decoration={markings, mark=at position 0.5 with {\arrow{>}}},
			postaction={decorate}
		},
		pairing/.style={dashed, thick} 
		]
		
		\def\radius{1.8}
		
		\foreach \i/\angle in {1/90, 2/0, 3/270, 4/180} {
			\node[tensor] (T\i) at (\angle:\radius) {$T$};
		}
		
		\foreach \i/\angle in {1/45, 2/315, 3/225, 4/135} {
			\node[op] (A\i) at (\angle:\radius) {$A_{\i}$};
		}
		
		\draw[arcarrow] (T1) to [bend left=15] (A1);
		\draw[arcarrow] (A1) to [bend left=15] (T2);
		\draw[arcarrow] (T2) to [bend left=15] (A2);
		\draw[arcarrow] (A2) to [bend left=15] (T3);
		\draw[arcarrow] (T3) to [bend left=15] (A3);
		\draw[arcarrow] (A3) to [bend left=15] (T4);
		\draw[arcarrow] (T4) to [bend left=15] (A4);
		\draw[arcarrow] (A4) to [bend left=15] (T1);
		
		\begin{scope}[pairing, blue]
			\draw (T1) to [bend right=30] (T2);
			\draw (T3) to [bend right=30] (T4);
		\end{scope}

		\node[below, align=left, font=\scriptsize] at (0, -\radius-1) {
			\textcolor{blue}{---(a) Pairing 1 (Adjacent)} 
		};
		
	\end{tikzpicture}
	\begin{tikzpicture}[
		node distance=2cm,
		tensor/.style={circle, draw=black, thick, fill=white, inner sep=2pt, minimum size=8mm},
		op/.style={rectangle, draw=black, thick, fill=white, inner sep=2pt, minimum size=8mm},
		arcarrow/.style={
			thick,
			decoration={markings, mark=at position 0.5 with {\arrow{>}}},
			postaction={decorate}
		},
		pairing/.style={dashed, thick} 
		]
		
		\def\radius{1.8}
		
		\foreach \i/\angle in {1/90, 2/0, 3/270, 4/180} {
			\node[tensor] (T\i) at (\angle:\radius) {$T$};
		}
		
			\foreach \i/\angle in {1/45, 2/315, 3/225, 4/135} {
			\node[op] (A\i) at (\angle:\radius) {$A_{\i}$};
		}
		
		\draw[arcarrow] (T1) to [bend left=15] (A1);
		\draw[arcarrow] (A1) to [bend left=15] (T2);
		\draw[arcarrow] (T2) to [bend left=15] (A2);
		\draw[arcarrow] (A2) to [bend left=15] (T3);
		\draw[arcarrow] (T3) to [bend left=15] (A3);
		\draw[arcarrow] (A3) to [bend left=15] (T4);
		\draw[arcarrow] (T4) to [bend left=15] (A4);
		\draw[arcarrow] (A4) to [bend left=15] (T1);
		

		\begin{scope}[pairing, orange]
			\draw (T1) to [bend left=45] (T4);
			\draw (T2) to [bend right=45] (T3);
		\end{scope}
		
		\node[below, align=left, font=\scriptsize] at (0, -\radius-1) {
			\textcolor{orange}{---(b) Pairing 2 (Alternate)}
		};
		
	\end{tikzpicture}
\begin{tikzpicture}[
node distance=2cm,
tensor/.style={circle, draw=black, thick, fill=white, inner sep=2pt, minimum size=8mm},
op/.style={rectangle, draw=black, thick, fill=white, inner sep=2pt, minimum size=8mm},
arcarrow/.style={
	thick,
	decoration={markings, mark=at position 0.5 with {\arrow{>}}},
	postaction={decorate}
},
pairing/.style={dashed, thick} 
]

\def\radius{1.8}

\foreach \i/\angle in {1/90, 2/0, 3/270, 4/180} {
	\node[tensor] (T\i) at (\angle:\radius) {$T$};
}

\foreach \i/\angle in {1/45, 2/315, 3/225, 4/135} {
	\node[op] (A\i) at (\angle:\radius) {$A_{\i}$};
}

\draw[arcarrow] (T1) to [bend left=15] (A1);
\draw[arcarrow] (A1) to [bend left=15] (T2);
\draw[arcarrow] (T2) to [bend left=15] (A2);
\draw[arcarrow] (A2) to [bend left=15] (T3);
\draw[arcarrow] (T3) to [bend left=15] (A3);
\draw[arcarrow] (A3) to [bend left=15] (T4);
\draw[arcarrow] (T4) to [bend left=15] (A4);
\draw[arcarrow] (A4) to [bend left=15] (T1);


\begin{scope}[pairing, green!70!black]
	\draw (T1) to  (T3);
	\draw (T2) to  (T4);
\end{scope}

\node[below, align=left, font=\scriptsize] at (0, -\radius-1) {
	\textcolor{green!70!black}{---(c) Pairing 3 (Cross)} 
};

\end{tikzpicture}
\caption{Schematic of the three pairing types. We use directed lines to represent matrix indices: an arrow pointing \textbf{toward} an operator corresponds to a \textbf{row index}, while an arrow pointing \textbf{away from} an operator corresponds to a \textbf{column index}. The translation of a diagram into a product of traces is straightforward. Begin at any operator and follow the direction of the arrows. Upon encountering a $T$, jump to the paired $T$ along the dashed line, and then continue following the arrows until a closed loop is completed. This loop corresponds to a single trace, with the operators appearing in the order encountered along the path. Repeat this procedure until all operators in the diagram have been traversed.}
\label{fig:pairs}
\end{center}
\end{figure}
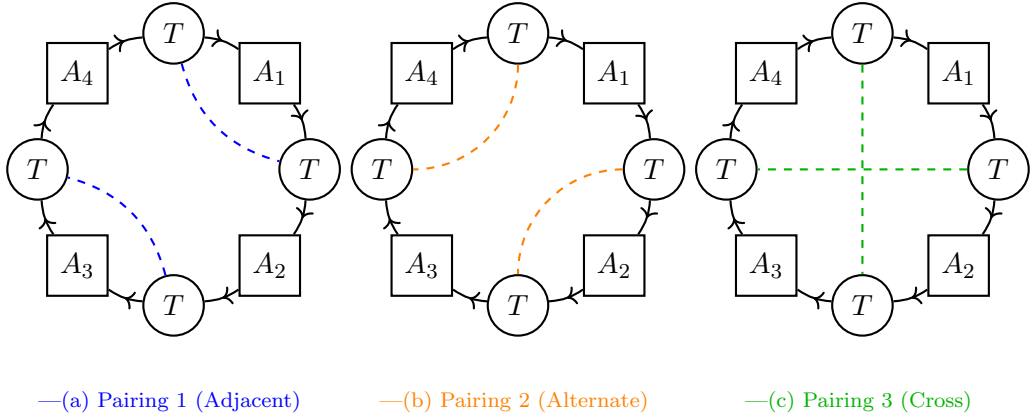
\subsection{Next-to-leading order perturbation}
To second order in $\mathcal{J}$, the contribution to the energy expectation value is
\begin{align}
\overline{E}_1^{(2)}(t)=\sum_{j+k=4}\mathbb{E}\left[\text{Tr}\left(U^{\dagger(j)}(t)H_{1}U^{(k)}(t)\rho_{init}\right)\right]\ed
\end{align}
We note the symmetry relation
\begin{align}
\mathbb{E}\left[\text{Tr}\left(U^{\dagger(j)}(t)H_{1}U^{(k)}(t)\rho_{init}\right)\right]=\mathbb{E}\left[\text{Tr}\left(U^{\dagger(k)}(t)H_{1}U^{(j)}(t)\rho_{init}\right)\right]^*.
\end{align}
Consequently, three distinct combinations contribute: $\{(j,k)\} = \{(0,4)+(4,0), (1,3)+(3,1), (2,2)\}$. As illustrated in Fig.~\ref{fig:pairs}, three distinct pairing patterns emerge. The evaluation requires computing ensemble averages of the form
\begin{equation}
\begin{aligned}
&\mathbb{E}(\text{Tr}(TA_1TA_2TA_3TA_4))\\
=&J^2 \left[\text{Tr}(A_1)\text{Tr}(A_3)\text{Tr}(A_2 A_4)+\text{Tr}(A_{2})\text{Tr}(A_{4})\text{Tr}(A_{1}A_{3})+\text{Tr}(A_4A_3A_2A_1)\right]\ed
\end{aligned}
\end{equation}
In the large-$N$ limit, the crossing pairing (Fig.~\ref{fig:pairs}c) is subleading, and we need only consider non-crossing (planar) contractions. After a lengthy but straightforward calculation (detailed in Appendix~\ref{appdix:NLorder}), we obtain 
\begin{equation}\label{eq:E1NL}
\begin{aligned}
\overline{E}_1^{(2)}(t)=&{\mathcal{J}^4 \over \mathcal{Z}}\int d[\epsilon]_{\beta_1\beta_2}d[\eta]d[\xi]\bigg[ \frac{\left(2\eta_{1}\left(\eta^{2}-\eta\xi+\xi^{2}+\epsilon^{2}-\epsilon(\eta+\xi)\right)-\epsilon_{1}(\eta-\xi)^{2}\right)\cos(t(\epsilon-\eta))}{(\eta-\xi)^{2}(\epsilon-\eta)^{2}(\epsilon-\xi)^{2}}\\&+\frac{\left(\epsilon_{1}(\eta-\xi)+2\eta_{1}(\epsilon-\eta)\right)\cos(t(\xi-\eta))}{(\eta-\xi)(\epsilon-\eta)^{2}(\epsilon-\xi)^{2}}-\frac{\left(2\eta_{1}(\epsilon-\eta)(\epsilon-\xi)+\epsilon_{1}(\eta-\xi)^{2}\right)\cos(t(\epsilon-\xi))}{(\eta-\xi)^{2}(\epsilon-\eta)^{2}(\epsilon-\xi)^{2}}\\&+\frac{t\left(\epsilon_{1}-2\eta_{1}\right)\sin(t(\epsilon-\eta))}{(\eta-\xi)(\epsilon-\eta)(\epsilon-\xi)}-\frac{t\epsilon_{1}\sin(t(\epsilon-\xi))}{(\eta-\xi)(\epsilon-\eta)(\epsilon-\xi)}+\frac{\epsilon_{1}(\eta-\xi)+2\eta_{1}(\xi-\epsilon)}{(\eta-\xi)(\epsilon-\eta)^{2}(\epsilon-\xi)^{2}}\bigg]\ed
\end{aligned}
\end{equation}
where we have used the shorthand $\epsilon = \epsilon_1+\epsilon_2$, $\eta = \eta_1+\eta_2$, and $\xi = \xi_1+\xi_2$. When energy denominators approach zero, L'H\^opital's rule should be applied to avoid numerical singularities. In numerical simulations, however, the contributions from configurations with vanishing denominators are suppressed by large $N$ in the sum, so these terms can be safely discarded. Alternatively, one can add a small imaginary part to the denominator and then take the real part of the entire expression.

The steady-state energy current at second order follows from taking the time derivative and applying
$\lim_{t\to \infty} {\sin(t x)\over x}= \pi \delta(x)$:
\begin{align}
\dot{\overline{E}}_1^{(2)}(\infty)=\frac{2\pi\mathcal{J}^4}{\mathcal{Z}}\left\{\int d[\epsilon]_{\beta_{1},\beta_{2}}d[\eta]d[\xi]\left[\frac{\left(\epsilon_{1}-\eta_{1}\right)\delta(\epsilon-\eta)}{(\epsilon-\xi)^{2}}+\frac{\left(\epsilon_{1}-\xi_{1}\right)\delta(\epsilon-\xi)}{(\epsilon-\eta)^{2}}\right]+\lim_{t\to\infty}I(t)/\pi\right\}\ed
\end{align}
where 
\begin{align}
I(t) = \int d[\epsilon]_{\beta_{1},\beta_{2}}d[\eta]d[\xi]\,
\frac{t\,(\epsilon_{1}-\eta_{1})}{(\eta-\xi)(\epsilon-\xi)}\cos(t(\epsilon-\eta))\ed
\label{eq:I_simplified}
\end{align}
Now observe that $t\cos(t(\epsilon-\eta)) = -\frac{\partial}{\partial\eta_{1}}\sin(t(\epsilon-\eta))$.
Insert this into~\eqref{eq:I_simplified} and integrate by parts in $\eta_{1}$:
\begin{equation}
\begin{aligned}
I(t) &= -\int d[\epsilon]_{\beta_{1},\beta_{2}}\rho_1(\eta_1)\rho_2(\eta_2)d\eta_{2}d[\xi] \int d\eta_{1}\,
\left(\frac{\partial}{\partial\eta_{1}}\sin(t(\epsilon-\eta))\right)
\frac{\epsilon_{1}-\eta_{1}}{(\eta-\xi)(\epsilon-\xi)} \\
&= \int d[\epsilon]_{\beta_{1},\beta_{2}}\rho_1(\eta_1)\rho_2(\eta_2)d\eta_{2}d[\xi] \int d\eta_{1}\,
\sin(t(\epsilon-\eta))\,
\frac{\partial}{\partial\eta_{1}} \left(\frac{\epsilon_{1}-\eta_{1}}{(\eta-\xi)(\epsilon-\xi)}\right)\co
\end{aligned}
\end{equation}
where the boundary term vanishes because the spectral densities $\rho_i$ contained in the measures decay sufficiently fast at infinity.
The derivative in the integrand is
\begin{align}
\frac{\partial}{\partial\eta_{1}} \left(\frac{\epsilon_{1}-\eta_{1}}{(\eta-\xi)(\epsilon-\xi)}\right)
= -\frac{1}{(\eta-\xi)(\epsilon-\xi)} - \frac{\epsilon_{1}-\eta_{1}}{(\eta-\xi)^2(\epsilon-\xi)}\ed
\end{align}
Under the usual regularity assumptions on the densities of states (smooth and rapidly decaying), each of the resulting terms is absolutely integrable.
The singularities at $\eta=\xi$ and $\epsilon=\xi$ are integrable in the principal-value sense, and the multiplication by the derivative only produces $(\eta-\xi)^{-2}$ type singularities, which remain integrable in three dimensions.
Hence the function
\begin{align}
F(\eta_{1},\dots) = \frac{\partial}{\partial\eta_{1}} \left(\frac{\epsilon_{1}-\eta_{1}}{(\eta-\xi)(\epsilon-\xi)}\right) \nonumber
\end{align}
belongs to $L^{1}$ with respect to all integration variables.
The Riemann--Lebesgue lemma then guarantees that the integral
\begin{align}
\int \rho_1(\eta_1)d\eta_{1}\, \sin(t(\epsilon-\eta))\,F(\eta_{1},\dots) \nonumber
\end{align}
vanishes as $t\to\infty$ for almost every value of the remaining variables, and dominated convergence implies that the full multidimensional integral tends to zero.
Consequently, $\lim_{t\to\infty}I(t)=0$.

Unlike the leading-order result, the denominators in the expression for \(\overline{E}_1^{(2)}\) cannot be eliminated by taking higher-order derivatives. This prevents a straightforward simplification of the integral in terms of Fourier transforms of the spectral functions. Consequently, even for the steady-state energy current, one generally faces a fivefold integral that is difficult to evaluate analytically. 
Nevertheless, the explicit second-order expression we derive is not only convenient for numerical evaluation but also amenable to analytical analysis. For instance, in the short-time regime, the integrand can be expanded to obtain corrections to the leading-order perturbative behavior.
\section{Examples and Numerical Simulations}
\label{sec:examples}
In the preceding section, we derived a general leading-order expression for the energy current. This expression exhibits universal features, including initial linear growth and eventual saturation to a constant value at long times. In this section, we illustrate these behaviors by evaluating the energy current explicitly for several concrete spectral densities. We denote the mean energy scales of $H_1$ and $H_2$ as $\mathcal{E}_1$ and $\mathcal{E}_2$, so that together with the coupling $\mathcal{J}$ we have three independent energy scales.

The specific models analyzed below are defined by the following densities of states:
\begin{itemize}
\item Gaussian model
\begin{align}
\rho_{\text{Gaussian}}\left(\epsilon\right)=\frac{e^{-\frac{\epsilon^{2}}{2\mathcal{E}^{2}}}}{\sqrt{2\pi\mathcal{E}^{2}}}\ed
\end{align}
\item Wigner semicircle distribution
\begin{align}
\rho_{\text{Semi-C}}\left(\epsilon\right)=\frac{1}{\mathcal{E}}\frac{\sqrt{4-\epsilon^{2}/\mathcal{E}^{2}}}{2\pi},-2\mathcal{E}\le\epsilon\le2 \mathcal{E} \ed
\end{align}
\item Constant distribution
\begin{align}
\rho_{\text{Const}}\left(\epsilon\right)=\frac{1}{4\mathcal{E}},-2\mathcal{E}\le\epsilon\le2\mathcal{E}\ed
\end{align}
\item Gamma distribution
\begin{align}
\rho_{\text{Gamma}}(\epsilon)=\frac{1}{\mathcal{E}}\frac{\left(\epsilon/\mathcal{E}\right)^{\alpha}e^{-\epsilon/\mathcal{E}}}{\Gamma(\alpha+1)},\epsilon\ge 0 \ed
\end{align}
For an $n$-dimensional gas, one has $\alpha = n/2-1$. 
\end{itemize}
Another noteworthy point is that, after taking the continuum limit, information about the spacing between adjacent energy levels is erased. To consider chaotic systems, one should take $H_1$ and $H_2$ as random matrix ensembles and then average the heat current over the ensemble. The result without ensemble averaging (i.e., ignoring spectral correlations) corresponds to integrable systems.
\subsection{Leading-order results}
As all expressions in this subsection are to leading order in \(\mathcal{J}\), we omit the superscript \(^{(1)}\) for brevity. The perturbative order remains unambiguous as it is explicitly indicated by the power of \(\mathcal{J}\).
\begin{figure}[ht]
\begin{center}
\includegraphics[width=0.48\textwidth]{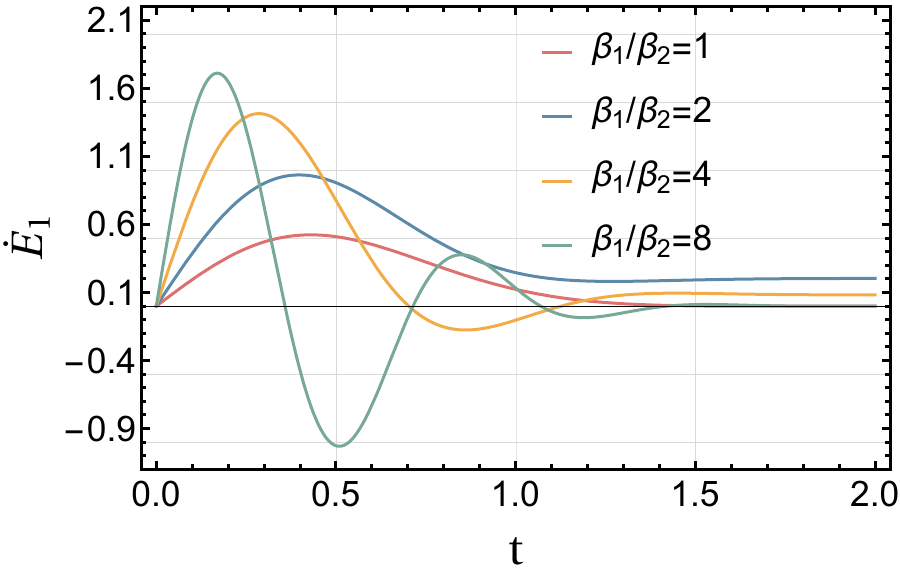}
\includegraphics[width=0.48\textwidth]{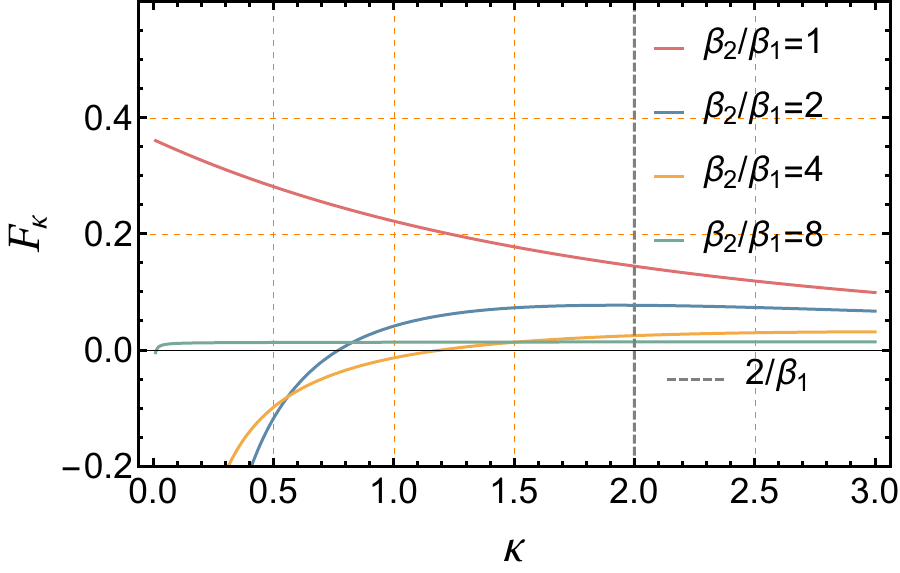}
\caption{Energy current and integrated energy flux for the Gaussian model with $\mathcal{J}=1$ and $\mathcal{E}=1$. We set $\beta_2=1$ in the left panel and $\beta_1=1$ in the right panel.}
\label{fig:current_Gaussian}
\end{center}
\end{figure} 
\begin{figure}[ht]
\begin{center}
\includegraphics[width=0.48\textwidth]{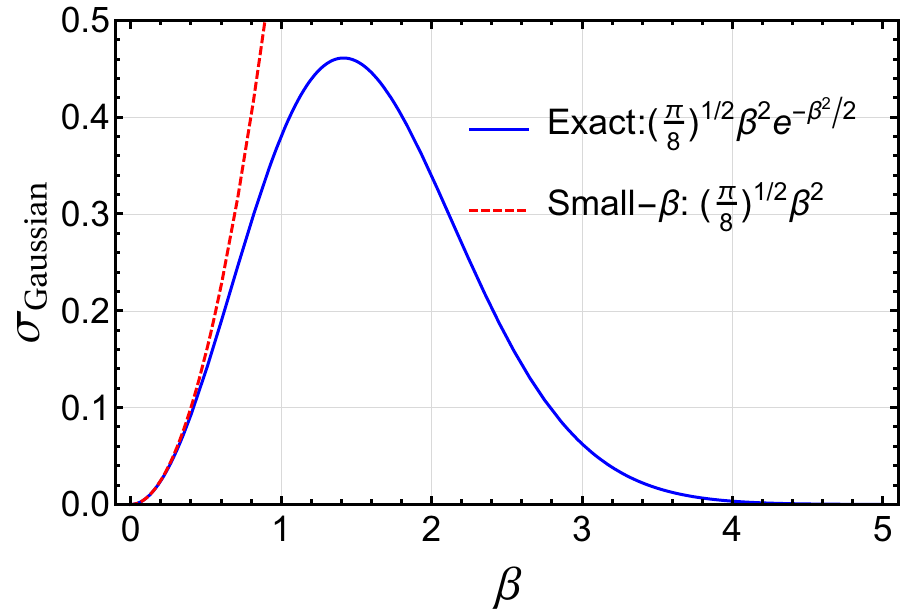}
\includegraphics[width=0.48\textwidth]{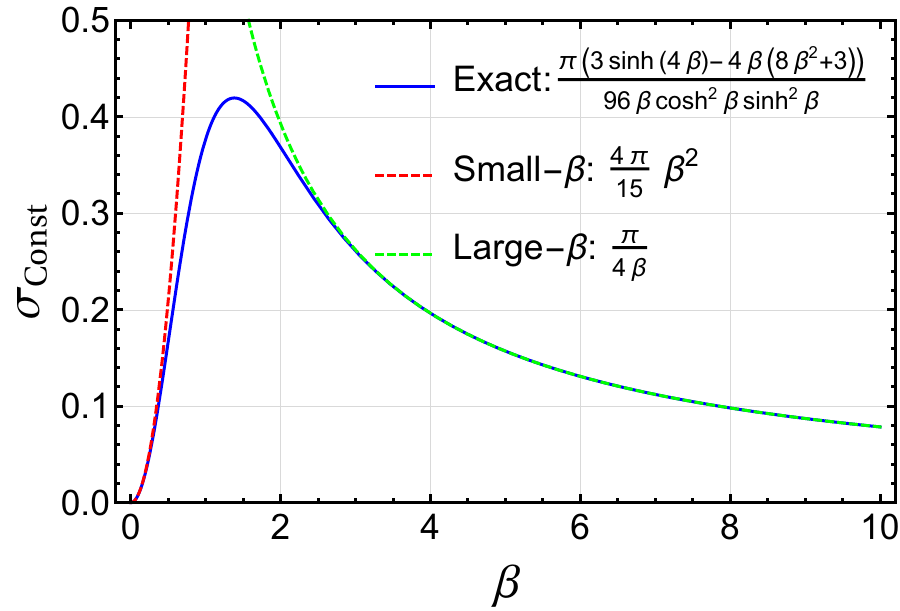}
\caption{Heat conductance for the Gaussian model and the constant distribution, with $\mathcal{J}=\mathcal{E}=1$.}
\label{fig:sigma}
\end{center}
\end{figure} 
One could consider two subsystems with different spectral distributions. For simplicity, however, we focus on the case in which both subsystems have the same type of spectrum and the same energy scale. Thus we take \(\rho_1 = \rho_2\) and write 
$\mathcal{E}_1=\mathcal{E}_2 =\mathcal{E}$. 
\paragraph{Gaussian model}
The Gaussian model is analytically tractable. From Eq.~\eqref{eq:ddot_E1}, we obtain the energy current
\begin{align}
\dot{\overline{E}}^{\text{Gaussian}}_1(t)&=\frac{\mathcal{J}^2}{2}\sqrt{\frac{\pi}{2}}\left(\beta_{1}-\beta_{2}\right)\mathcal{E}e^{-\frac{1}{8}\left(\beta_{1}+\beta_{2}\right){}^{2} \mathcal{E}^2}\Re\left[\text{erf}\left(\frac{ \mathcal{E}\left(4t-i\left(\beta_{1}+\beta_{2}\right)\right)}{2\sqrt{2}}\right)\right]\nn
&\newline +\mathcal{J}^2e^{-2\mathcal{E}^{2}t^{2}}\sin\left(\left(\beta_{1}+\beta_{2}\right)\mathcal{E}^{2}t\right)\ed
\end{align}
In the steady-state limit \(t \to \infty\), this simplifies to
\begin{align}
\dot{\overline{E}}^{\text{Gaussian}}_1(\infty)=\frac{\mathcal{J}^2}{2} \sqrt{\frac{\pi }{2}} \left(\beta _1-\beta _2\right) \mathcal{E} e^{-\frac{1}{8} \left(\beta _1+\beta _2\right){}^2
\mathcal{E}^2}\co
\end{align}
from which the heat conductance follows as
\begin{align}
\sigma_{\text{Gaussian}}=\frac{\mathcal{J}^2}{2}\sqrt{\frac{\pi}{2}}\beta^{2}\mathcal{E}e^{-\frac{1}{2}\beta{}^{2}\mathcal{E}^2}\ed
\end{align}
The integrated energy flux, defined in Eq.~\eqref{eq:integrable-flow}, is also computable:
\begin{align}
F_\kappa^{\text{Gaussian}}&=\sqrt{\frac{\pi}{2}}\frac{\mathcal{J}^2\left(\left(\beta_{1}-\beta_{2}\right)\mathcal{E}^{2}+i\kappa\right)}{4\kappa\mathcal{E}}e^{\frac{\left(\kappa+i\left(\beta_{1}+\beta_{2}\right)\mathcal{E}^{2}\right){}^{2}}{8\mathcal{E}^{2}}}\text{erfc}\left(\frac{\kappa+i\left(\beta_{1}+\beta_{2}\right)\mathcal{E}^{2}}{2\sqrt{2}\mathcal{E}}\right)+\text{h.c.}
\end{align}
where \(\text{erf}(z)\equiv \frac{2}{\sqrt{\pi}}\int_{0}^{z}e^{-t^{2}}dt\) is the error function, \(\text{erfc}(z) = 1 - \text{erf}(z)\) is the complementary error function, and ``\text{h.c.}'' denotes the complex-conjugate contribution. These results are illustrated in Fig.~\ref{fig:current_Gaussian} and in the left panel of Fig.~\ref{fig:sigma}. The Gaussian model exhibits oscillations in the energy current, which can be attributed to the fact that its spectrum is unbounded from below. In physical systems, however, the ground-state energy is finite. For the other spectral distributions considered below, these oscillations do not appear. Moreover, as shown in the figure, the integrated energy flux satisfies the inequality proposed in \cite{Almheiri:2019jqq}.
\paragraph{Constant distribution} For the constant distribution, the energy current is given by a more complicated integral expression
\begin{align}
\dot{\overline{E}}_{1}^{\text{Const}}(t)&=\frac{\beta_{1}\beta_{2}\mathcal{J}^{2}}{8\mathcal{E}^{2}\sinh\left(2\beta_{1}\mathcal{E}\right)\sinh\left(2\beta_{2}\mathcal{E}\right)}\int_{0}^{t}d\tau\frac{\sin(2\tau\mathcal{E})\sinh\left(2\mathcal{E}\left(\beta_{2}+i\tau\right)\right)}{\tau^{3}\left(\tau-i\beta_{1}\right){}^{2}\left(\tau-i\beta_{2}\right)}\times \nn
&\newline\bigg[4i\tau\mathcal{E}\left(\beta_{1}+i\tau\right)\sin\left(4\tau\mathcal{E}-2i\beta_{1}\mathcal{E}\right)+\left(-2\tau+i\beta_{1}\right)\cos\left(4\tau\mathcal{E}-2i\beta_{1}\mathcal{E}\right)\nn
&\newline+\left(2\tau-i\beta_{1}\right)\cosh\left(2\beta_{1}\mathcal{E}\right)\bigg]+\text{h.c.}
\end{align}
In the special case where \(\beta_1 = \beta_2 = \beta\), this reduces to a closed form
\begin{align}
\dot{\overline{E}}_{1}^{\text{Const}}(t)=\frac{\beta^{2}\mathcal{J}^{2}\sin^{2}(2t\mathcal{E})\left(\left(\beta^2-t^{2}\right)\sin(4t\mathcal{E})\sinh(4\beta\mathcal{E})+2\beta t-2\beta t\cos(4t\mathcal{E})\cosh(4\beta\mathcal{E})\right)}{8t^{2}\mathcal{E}^{2}\left(\beta^{2}+t^{2}\right)^{2}\sinh^{2}(2\beta\mathcal{E})}\co
\end{align}
The steady-state current is evaluated using contour integration (see Appendix~\ref{appdix:derive}), yielding
\begin{equation}
\begin{aligned}\label{eq:current_const}
	\dot{\overline{E}}_{1}^{(1)}(\infty)=&\frac{\mathcal{J}^{2}\text{csch}\left(2\beta_{1}\mathcal{E}\right)\text{csch}\left(2\beta_{2}\mathcal{E}\right)}{32\mathcal{E}^{2}}\bigg[-8\pi\mathcal{E}^{2}e^{-2\left(\beta_{1}+\beta_{2}\right)\mathcal{E}}\left(e^{4\beta_{1}\mathcal{E}}-e^{4\beta_{2}\mathcal{E}}\right)\\&-\frac{4\pi\left(\beta_{1}^{2}-\beta_{2}\beta_{1}+\beta_{2}^{2}\right)\mathcal{E}e^{-2\left(\beta_{1}+\beta_{2}\right)\mathcal{E}}\left(e^{4\beta_{1}\mathcal{E}}+e^{4\beta_{2}\mathcal{E}}\right)}{\beta_{1}\left(\beta_{1}-\beta_{2}\right)\beta_{2}}\\&+\frac{\pi e^{-2\left(\beta_{1}+\beta_{2}\right)\mathcal{E}}\left(\left(\beta_{1}^{4}-2\beta_{2}\beta_{1}^{3}\right)\left(e^{4\beta_{1}\mathcal{E}}+1\right)\left(e^{4\beta_{2}\mathcal{E}}-1\right)+\left(2\beta_{2}^{3}\beta_{1}-\beta_{2}^{4}\right)\left(e^{4\beta_{1}\mathcal{E}}-1\right)\left(e^{4\beta_{2}\mathcal{E}}+1\right)\right)}{\beta_{1}^{2}\left(\beta_{1}-\beta_{2}\right){}^{2}\beta_{2}^{2}}\ed
\end{aligned}
\end{equation}
The energy conductance is obtained by expanding the above expression for \(\beta_{1,2} = \beta \pm \delta\beta/2\) to first order in \(\delta\beta\):
%
\begin{align}
\sigma_\text{Const}
&=\frac{\pi\mathcal{J}^{2}\left(3\sinh\left(4\beta\mathcal{E}\right)-4\beta\mathcal{E}\left(8\beta^{2}\mathcal{E}^{2}+3\right)\right)}{96\beta\mathcal{E}^{2}\cosh^{2}\left(\beta\mathcal{E}\right)\sinh^{2}\left(\beta\mathcal{E}\right)}\co
\end{align}
with the asymptotic limits
\begin{align}
\sigma_{\beta\mathcal{E}\ll1}\sim\frac{4\pi\mathcal{J}^{2}\mathcal{E}}{15}\beta^{2},\sigma_{\beta\mathcal{E}\gg1}\sim\frac{\pi\mathcal{J}^{2}}{4\beta\mathcal{E}^{2}}\ed
\end{align}
These results are plotted in Fig.~\ref{fig:sigma}.
\paragraph{Semicircle distribution}
Wigner's semicircle law is the universal spectral density for large Gaussian random matrices and provides a paradigmatic example of a distribution with compact support.  
We take the density of states for both subsystems to be
\begin{align}
\rho_{\text{Semi-C}}(\epsilon)=\frac{1}{\mathcal{E}}\frac{\sqrt{4-\epsilon^{2}/\mathcal{E}^{2}}}{2\pi},
\qquad -2\mathcal{E}\le\epsilon\le2\mathcal{E},
\label{eq:rho_semi}
\end{align}
so that the bandwidth is \(4\mathcal{E}\) and the normalization is \(\int_{-2\mathcal{E}}^{2\mathcal{E}}\rho_{\text{Semi-C}}(\epsilon)\,d\epsilon=1\).
Because the spectrum is bounded, energy transfer cannot access arbitrarily high frequencies, which regularizes the long-time behavior and eliminates the artificial oscillations that appear for unbounded densities such as the Gaussian model.
Physically, a bounded spectrum implies a well-defined ground state and a finite maximal excitation energy.

To obtain the time-dependent energy change rate, we insert the semicircle density~\eqref{eq:rho_semi} into the general leading-order result~\eqref{eq:ddot_E1}.
The required integrals can be expressed as
\begin{align}
\dot{\overline{E}}_{1}^{\text{Semi-C}}(t)=&\int_{0}^{t}d\tau\frac{2\beta_{1}\beta_{2}\mathcal{J}^{2}J_{1}(2\mathcal{E}\tau)I_{1}\left(2\mathcal{E}\left(i\tau+\beta_{2}\right)\right)}{\tau^{2}\mathcal{E}\left(\beta_{1}+i\tau\right)\left(\tau-i\beta_{2}\right)I_{1}\left(2\mathcal{E}\beta_{1}\right)I_{1}\left(2\mathcal{E}\beta_{2}\right)}\nn&\times\left(J_{2}(2\mathcal{E}\tau)I_{1}\left(2\mathcal{E}\left(i\tau+\beta_{1}\right)\right)-iJ_{1}(2\mathcal{E}\tau)I_{2}\left(2\mathcal{E}\left(i\tau+\beta_{1}\right)\right)\right)+\text{h.c.}
\end{align}
Here \(J_1\) is the Bessel function of the first kind, and \(I_1,\,I_2\) are modified Bessel functions of the first kind.
For early times \(t\ll \mathcal{E}^{-1}\) one recovers the universal linear growth \(\dot{\overline{E}}_1(t)\propto\mathcal{J}^{2}\mathcal{E}t\), in agreement with the short-time expansion discussed in Sec.~\ref{sec:LvNL}.

The long-time steady-state current follows from the general formula~\eqref{eq:current_infty_w} after substituting the Fourier transform of the semicircle density.
The result is a single integral over the frequency \(\omega\):
\begin{align}
&\dot{\overline{E}}_{1}^{\text{Semi-C}}(\infty)=\int_{-\infty}^{+\infty}d\omega\frac{\beta_{1}\beta_{2}\mathcal{J}^{2}J_{1}(2\mathcal{E}\omega){}^{2}}{\mathcal{E}\omega^{2}\left(\beta_{1}-i\omega\right)\left(\beta_{2}-i\omega\right)I_{1}\left(2\mathcal{E}\beta_{1}\right)I_{1}\left(2\mathcal{E}\beta_{2}\right)}\nn&\newline\times\left(I_{1}\left(2\mathcal{E}\left(\beta_{2}-i\omega\right)\right)I_{2}\left(2\mathcal{E}\left(\beta_{1}-i\omega\right)\right)-I_{1}\left(2\mathcal{E}\left(\beta_{1}-i\omega\right)\right)I_{2}\left(2\mathcal{E}\left(\beta_{2}-i\omega\right)\right)\right)\ed
\end{align}
The antisymmetry under \(\beta_1\leftrightarrow\beta_2\) reflects the fact that, on average, energy flows from the hotter subsystem to the colder one.
The integrand decays as \(|\omega|^{-3}\) at large \(|\omega|\), guaranteeing convergence, and the combination of Bessel functions encodes the overlap of the spectral densities and the thermal occupation factors.

From the steady-state current, we extract the heat conductance by specializing the general expression~\eqref{eq:sigma_w} to the semicircle density and expanding in a small temperature difference.
Setting \(\beta_1=\beta-\delta\beta/2,\;\beta_2=\beta+\delta\beta/2\) and keeping the linear term in \(\delta\beta\) gives
\begin{equation}
\begin{aligned}
	\sigma_{\text{Semi-C}}&=\int d\omega\frac{\beta^{4}\mathcal{J}^{2}J_{1}(2\mathcal{E}\omega){}^{2}}{\mathcal{E}\omega^{2}(\beta-i\omega)^{3}I_{1}(2\mathcal{E}\beta){}^{2}}\bigg[-2\mathcal{E}(\beta-i\omega)I_{2}(2\mathcal{E}(\beta-i\omega)){}^{2}\\&\newline+I_{1}(2\mathcal{E}(\beta-i\omega))(I_{2}(2\mathcal{E}(\beta-i\omega))+2\mathcal{E}(\beta-i\omega)I_{3}(2\mathcal{E}(\beta-i\omega)))\bigg]\ed
\end{aligned}
\end{equation}
The integral can be evaluated numerically for any finite \(\beta\mathcal{E}\).
Its asymptotic forms are
\begin{align}
\sigma_{\text{Semi-C}} \sim 
\begin{cases}
	c_1\,\mathcal{J}^2\mathcal{E}\,\beta^{2}, & \beta\mathcal{E}\ll 1,\\[4pt]
	c_2\,\dfrac{\mathcal{J}^2}{\beta^2\,\mathcal{E}^{3}}, & \beta\mathcal{E}\gg 1,
\end{cases}
\label{eq:sigma_semi_asymp}
\end{align}
with positive constants \(c_1\) and \(c_2\).

As mentioned earlier, when $H_1$ and $H_2$ describe chaotic systems, we may model them by GUE random matrices. In that case, the heat current should be averaged over the spectral ensemble. As an approximation, we implement this averaging in Eq.~\eref{eq:dE1J1} by the replacement
\begin{equation}
\begin{aligned}
	&\rho(\epsilon_{1})\rho(\epsilon_{2})\rho(\widetilde{\epsilon}_{1})\rho(\widetilde{\epsilon}_{2})\to\rho_{\text{Semi-C}}(\epsilon_{1})\rho_{\text{Semi-C}}(\epsilon_{2})\rho_{\text{Semi-C}}(\widetilde{\epsilon}_{1})\rho_{\text{Semi-C}}(\widetilde{\epsilon}_{2})\\&\newline+R_{2}(\epsilon_{1},\epsilon_{2})\rho_{\text{Semi-C}}(\widetilde{\epsilon}_{1})\rho_{\text{Semi-C}}(\widetilde{\epsilon}_{2})+R_{2}(\epsilon_{1},\widetilde{\epsilon}_{2})\rho_{\text{Semi-C}}(\widetilde{\epsilon}_{1})\rho_{\text{Semi-C}}(\epsilon_{2})+\ldots
\end{aligned}
\end{equation}
where $R_{2}(\epsilon_{1},\epsilon_{2})\equiv\mathbb{E}\left(\left(\rho(\epsilon_{1})-\overline{\rho}(\epsilon_{1})\right)\left(\rho(\epsilon_{2})-\overline{\rho}(\epsilon_{2})\right)\right)$ is the connected pair correlation function \cite{MEHTA1960395,GAUDIN1961447,Dyson1962}:
\begin{equation}
\begin{aligned}
	R_{2}(\epsilon_{1},\epsilon_{2})=-\frac{\sin^{2}\left[N(\epsilon_{1}-\epsilon_{2})/\mathcal{E}\right]}{\left[\pi N(\epsilon_{1}-\epsilon_{2})\right]^{2}}+\frac{1}{N\mathcal{E}\pi}\delta(\epsilon_{1}-\epsilon_{2})\ed
\end{aligned}
\end{equation}
The connected correlator $R_2$ plays an important role in the spectral form factor (SFF) of the GUE \cite{Cotler:2016fpe}. In the ramp regime $1\ll t<N/\mathcal{E}$, the contribution of spectral correlations to the SFF exceeds that of the averaged spectral density, leading to a ramp that grows linearly in time. This ramp is a well-known diagnostic of quantum chaos and reflects level repulsion. In the present heat-current calculation, however, the prethermalization plateau has a duration that scales as $\sim N$, while the contribution of $R_2$ to the heat current is suppressed by powers of $N$. Compared with the $O(1)$ contribution from the averaged spectral density throughout the plateau, this correction is negligible. Our result therefore indicates that, in the large-$N$ limit, random coupling washes out the distinction between chaotic and integrable spectra at the level of the averaged heat current.
\begin{figure}[ht]
\begin{center}
	\includegraphics[width=0.48\textwidth]{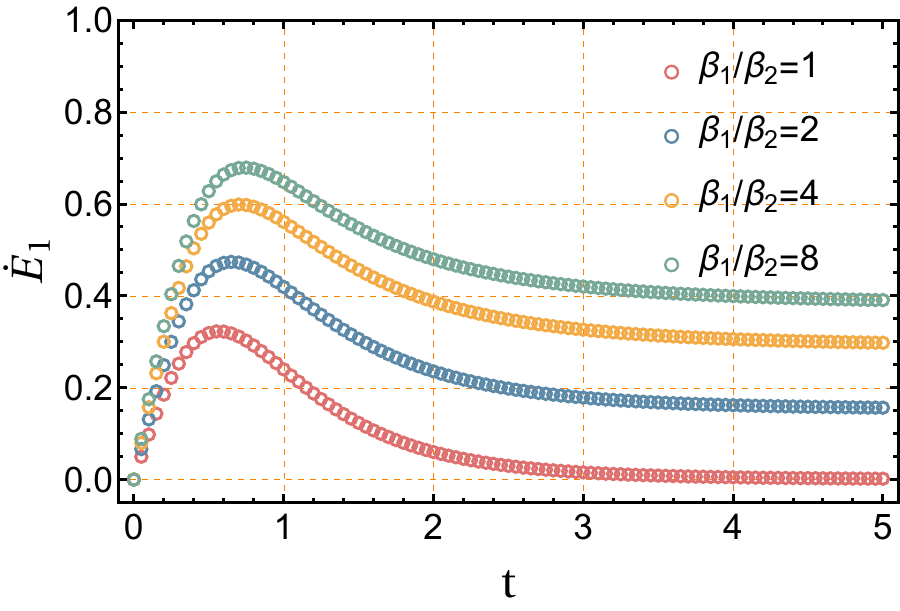}
	\includegraphics[width=0.48\textwidth]{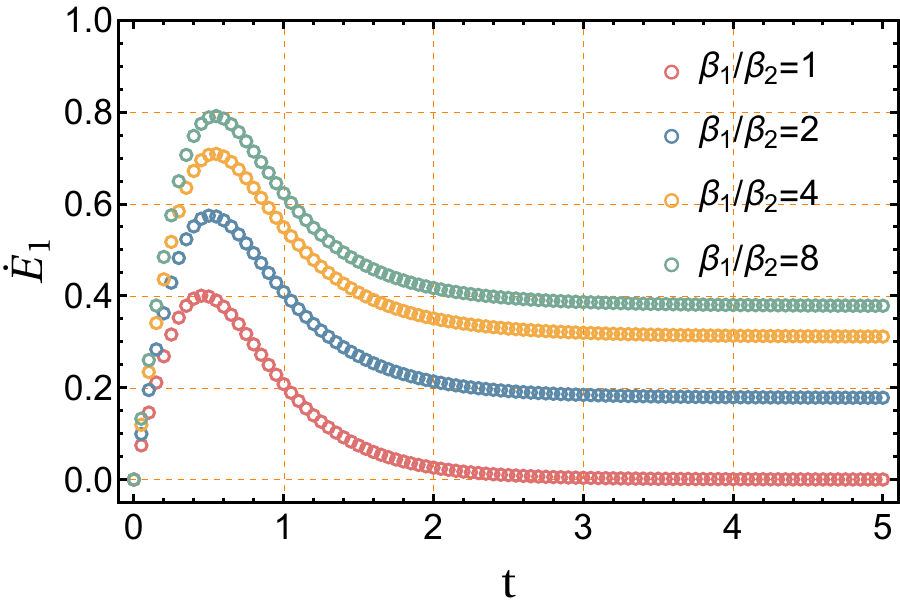}
	\caption{Energy current for the Gamma model with $\mathcal{J}=\mathcal{E}=\beta_2=1$. \textbf{Left}: $\alpha=0$, corresponding to a two-dimensional gas. \textbf{Right}: $\alpha=1/2$, corresponding to a three-dimensional gas.}
	\label{fig:current_Gamma}
\end{center}
\end{figure}

\paragraph{Gamma distribution}
The Gamma distribution provides a semi-infinite density of states that vanishes as a power law near the lower edge and decays exponentially at high energy.  
It is defined by
\begin{align}
\rho_{\text{Gamma}}(\epsilon)=\frac{1}{\mathcal{E}}\frac{\left(\epsilon/\mathcal{E}\right)^{\alpha}e^{-\epsilon/\mathcal{E}}}{\Gamma(\alpha+1)},
\qquad \epsilon\ge 0
\label{eq:rho_gamma}
\end{align}
where \(\alpha>-1\) is a shape parameter and \(\mathcal{E}\) sets the overall energy scale.  
For an \(n\)-dimensional ideal gas one has \(\alpha = n/2-1\); in particular \(\alpha=0\) corresponds to a two-dimensional gas and \(\alpha=1/2\) to a three-dimensional gas.  
The parameter \(\alpha\) controls the low-energy behavior of the density of states and, in turn, the low-temperature dependence of transport.

Substituting the Gamma density~\eqref{eq:rho_gamma} into the general leading-order result~\eqref{eq:dE1J1} and performing one of the spectral integrals yields the time-dependent energy change rate as a single integral over a time variable,
\begin{align}
\dot{\overline{E}}_1^{\text{Gamma}}(t) &=
\int_{0}^{t} d\tau\,
\frac{(\alpha+1)\mathcal{J}^{2}\mathcal{E}^{2}
	(\beta_{1}+2i\tau)
	(\beta_{1}\mathcal{E}+1)^{\alpha+1}
	(\beta_{2}\mathcal{E}+1)^{\alpha+1}}
{i(\tau\mathcal{E}+i)^{3}
	(1-i\tau\mathcal{E})^{2\alpha}
	(\beta_{1}\mathcal{E}+i\tau\mathcal{E}+1)^{\alpha+2}
	(\beta_{2}\mathcal{E}+i\tau\mathcal{E}+1)^{\alpha+1}}
+ \text{h.c.}
\label{eq:dE1_Gamma_t}
\end{align}
For early times \(t\ll \mathcal{E}^{-1}\), one recovers the universal linear growth \(\dot{\overline{E}}_1(t)\propto\mathcal{J}^{2}t\), while at later times the decay of the integrand guarantees convergence to a steady value.

In the long-time limit the frequency-space formula~\eqref{eq:current_infty_w} gives the steady-state current
\begin{align}
\dot{\overline{E}}_1^{\text{Gamma}}(\infty) &=
\int_{-\infty}^{+\infty} d\omega\,
\frac{(\alpha+1)(\beta_{2}-\beta_{1})\mathcal{J}^{2}\mathcal{E}^{2}
	(\beta_{1}\mathcal{E}+1)^{\alpha+1}
	(\beta_{2}\mathcal{E}+1)^{\alpha+1}}
{2(\mathcal{E}\omega-i)^{2}
	(1+i\mathcal{E}\omega)^{2\alpha}
	\bigl(1+\mathcal{E}(\beta_{1}-i\omega)\bigr)^{\alpha+2}
	\bigl(1+\mathcal{E}(\beta_{2}-i\omega)\bigr)^{\alpha+2}}.
\label{eq:dE1_Gamma_infty}
\end{align}
The antisymmetry under \(\beta_{1}\leftrightarrow\beta_{2}\) is manifest in the prefactor \((\beta_{2}-\beta_{1})\), showing that energy flows from the hotter subsystem to the colder one on average.

Expanding the steady-state current for a small temperature difference \(\beta_{1,2}=\beta\pm\delta\beta/2\) and applying the general expression~\eqref{eq:sigma_w} for the conductance, one can evaluate the integral analytically.  
The result is
\begin{align}
\sigma_{\text{Gamma}} &=
\frac{\pi(\alpha+1)\,\beta^{2}\,\mathcal{J}^{2}\,\mathcal{E}\,
	\Gamma(4\alpha+5)\,
	(\beta\mathcal{E}+1)^{2\alpha+2}}
{\Gamma(2\alpha+2)\,\Gamma(2\alpha+4)\,
	(\beta\mathcal{E}+2)^{4\alpha+5}}.
\label{eq:sigma_Gamma}
\end{align}
The two limiting behaviors of the conductance are
\begin{align}
\sigma_{\text{Gamma}} \sim
\begin{cases}
	\displaystyle
	\frac{\pi\,2^{-4\alpha-5}\,(\alpha+1)\,\beta^{2}\,\mathcal{J}^{2}\,\mathcal{E}\,
		\Gamma(4\alpha+5)}
	{\Gamma(2\alpha+2)\,\Gamma(2\alpha+4)},
	& \beta\mathcal{E}\ll 1,\\[12pt]
	\displaystyle
	\frac{\pi\,(\alpha+1)\,\beta\,\mathcal{J}^{2}\,
		\Gamma(4\alpha+5)\,
		(\beta\mathcal{E})^{-2(\alpha+1)}}
	{\Gamma(2\alpha+2)\,\Gamma(2\alpha+4)},
	& \beta\mathcal{E}\gg 1.
\end{cases}
\label{eq:sigma_Gamma_asymp}
\end{align}

In the calculations presented above, we have considered only gapless systems. In many quantum many-body systems, however, an energy gap separates the ground state from the excited states. We therefore examine this case by replacing the density of states as 
\begin{align}
\rho(\epsilon)\to\frac{1}{N}\delta(\epsilon-\epsilon_{g})+\frac{N-1}{N}\rho(\epsilon)\ed
\end{align}
As in the earlier discussion of chaotic systems, the corrections arising from the energy gap are suppressed at large $N$ and can therefore be neglected throughout the prethermalization process.
\subsection{Numerical simulations}
\begin{figure}[htbp]
\begin{center}
	\includegraphics[width=0.48\textwidth]{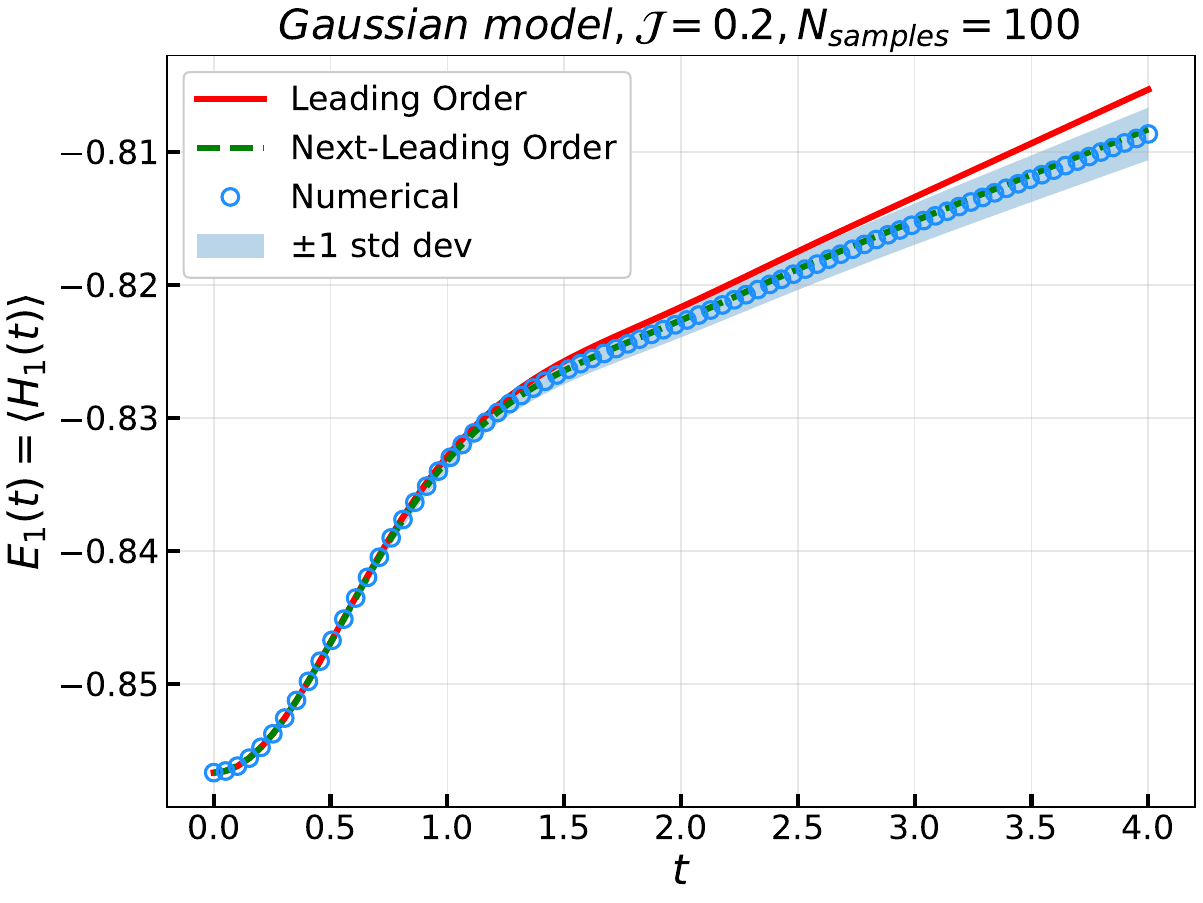}
	\includegraphics[width=0.48\textwidth]{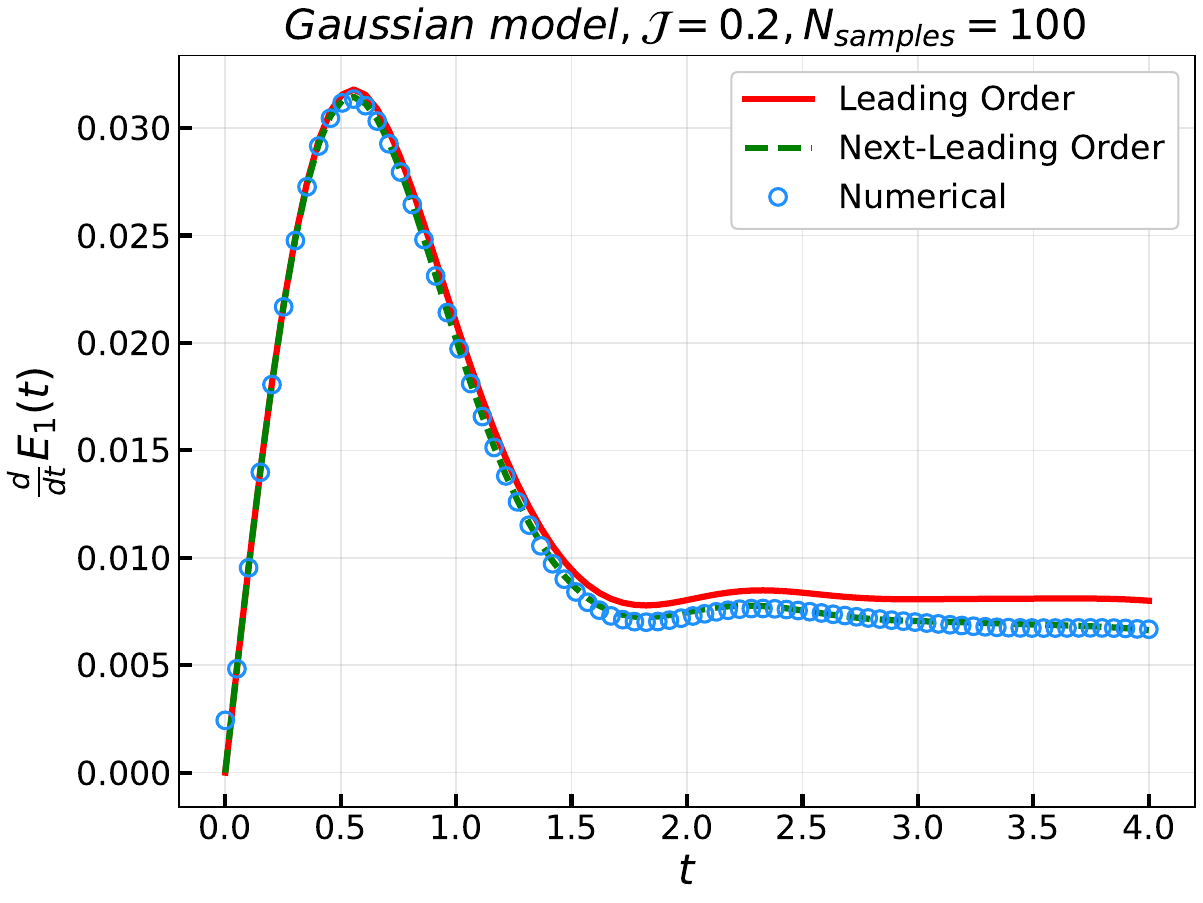}
	\includegraphics[width=0.48\textwidth]{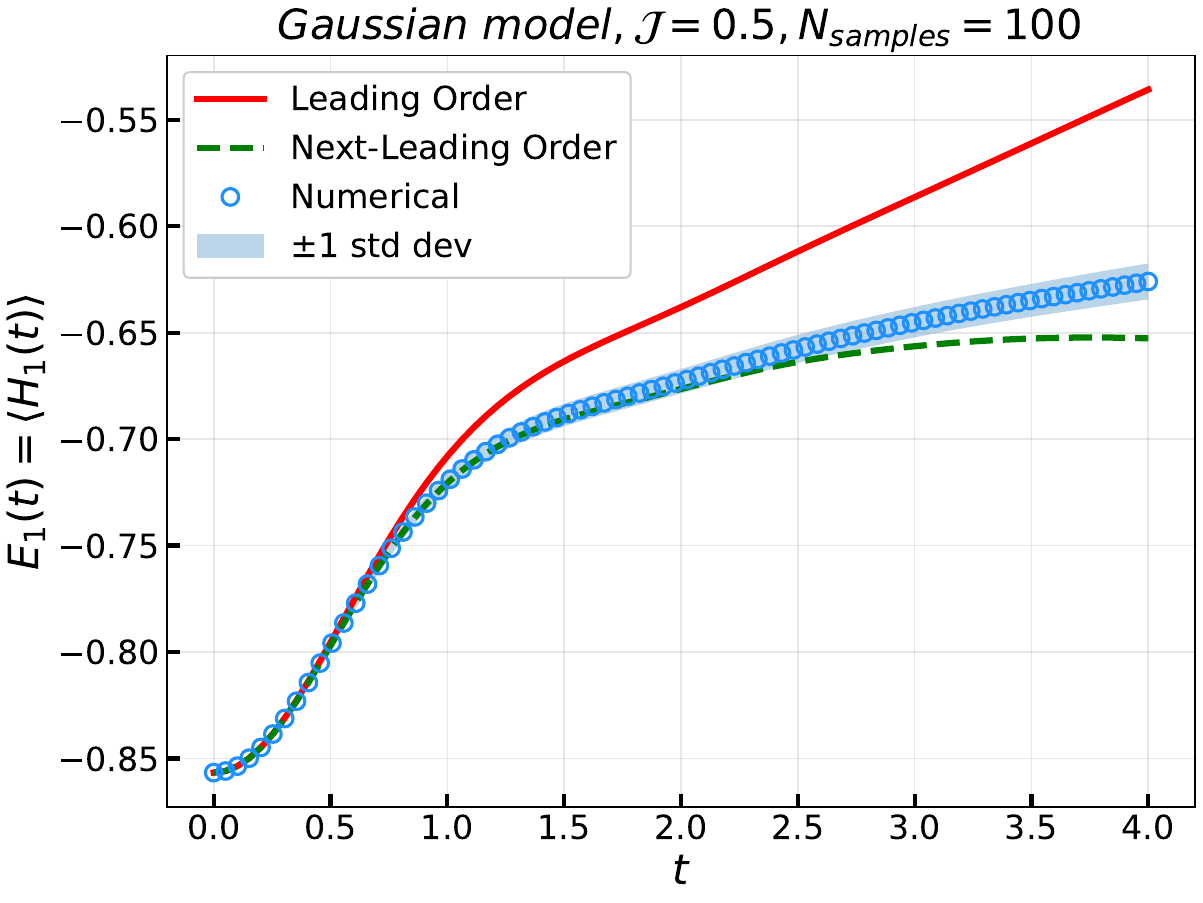}
	\includegraphics[width=0.48\textwidth]{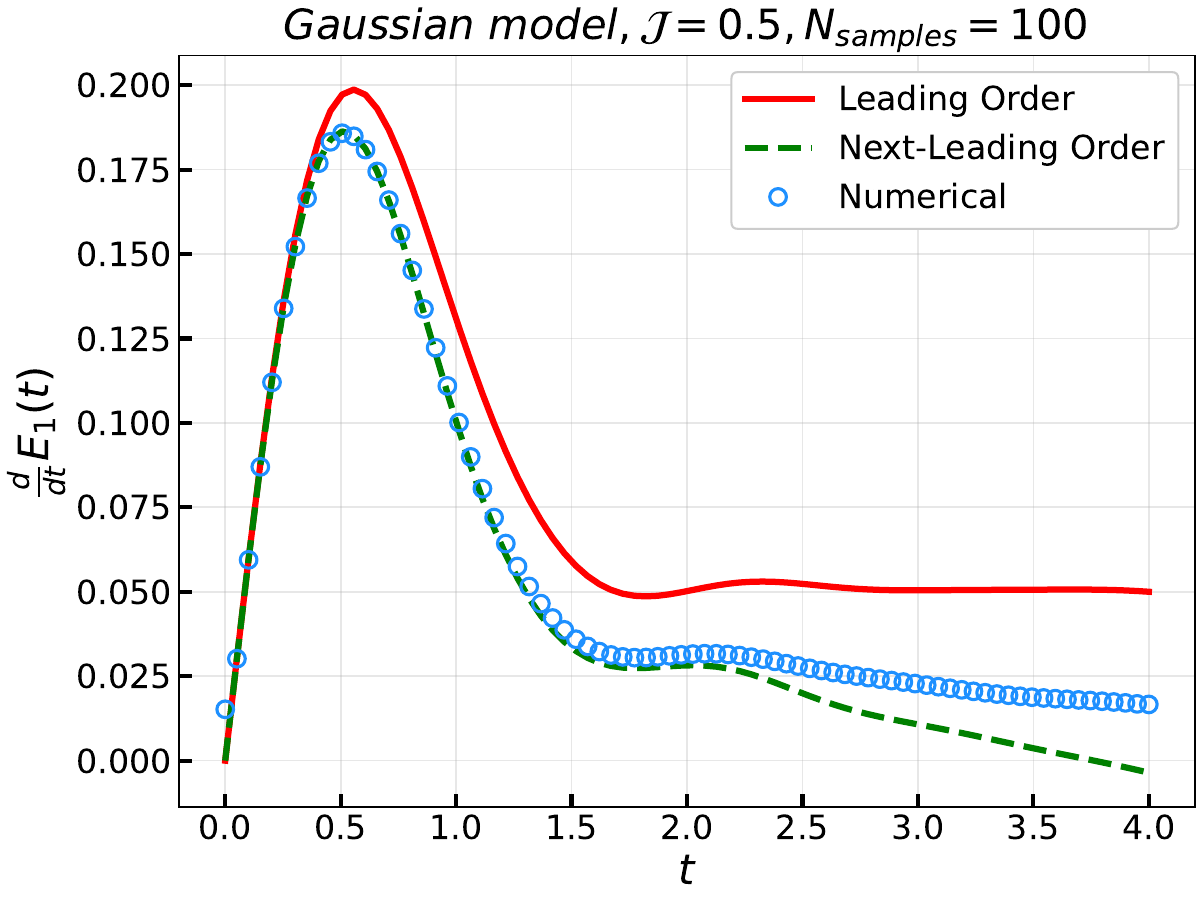}
	\includegraphics[width=0.48\textwidth]{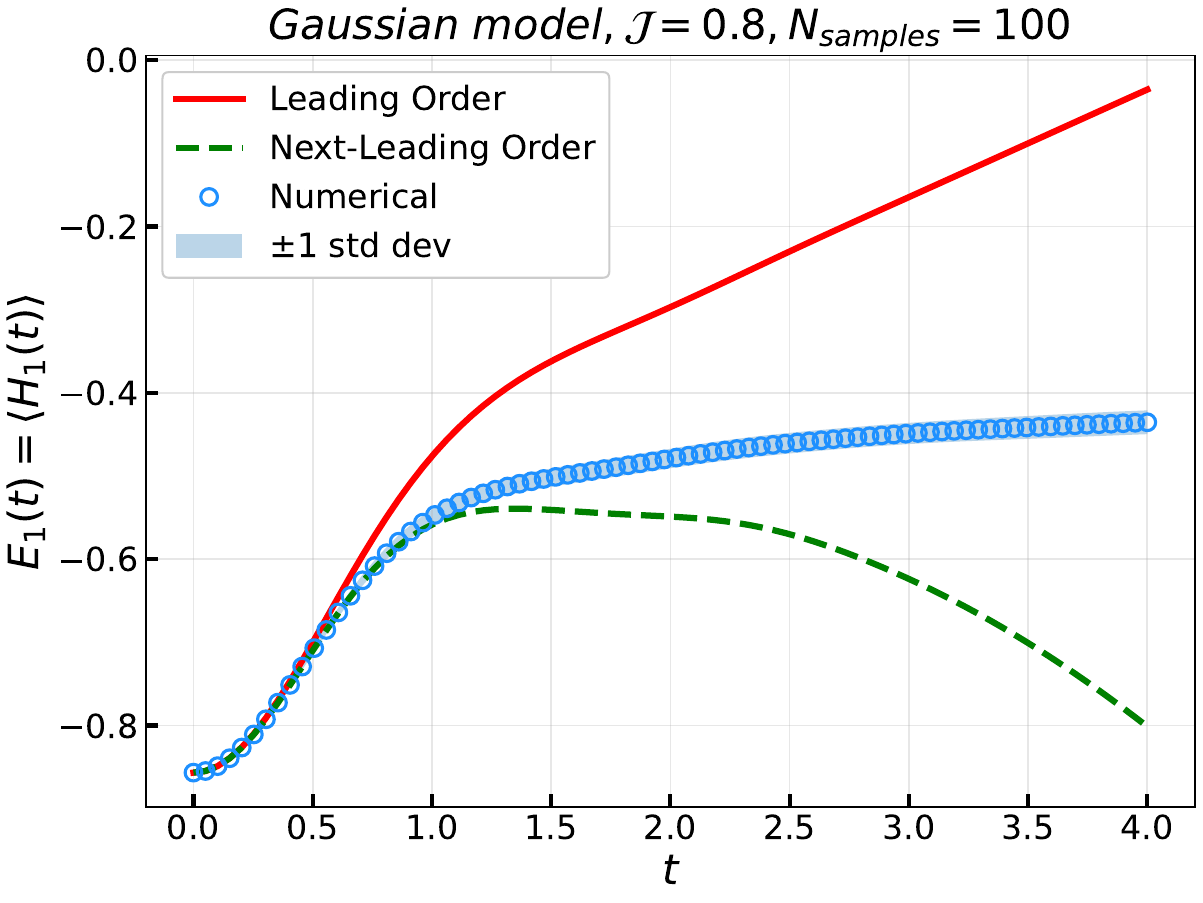}
	\includegraphics[width=0.48\textwidth]{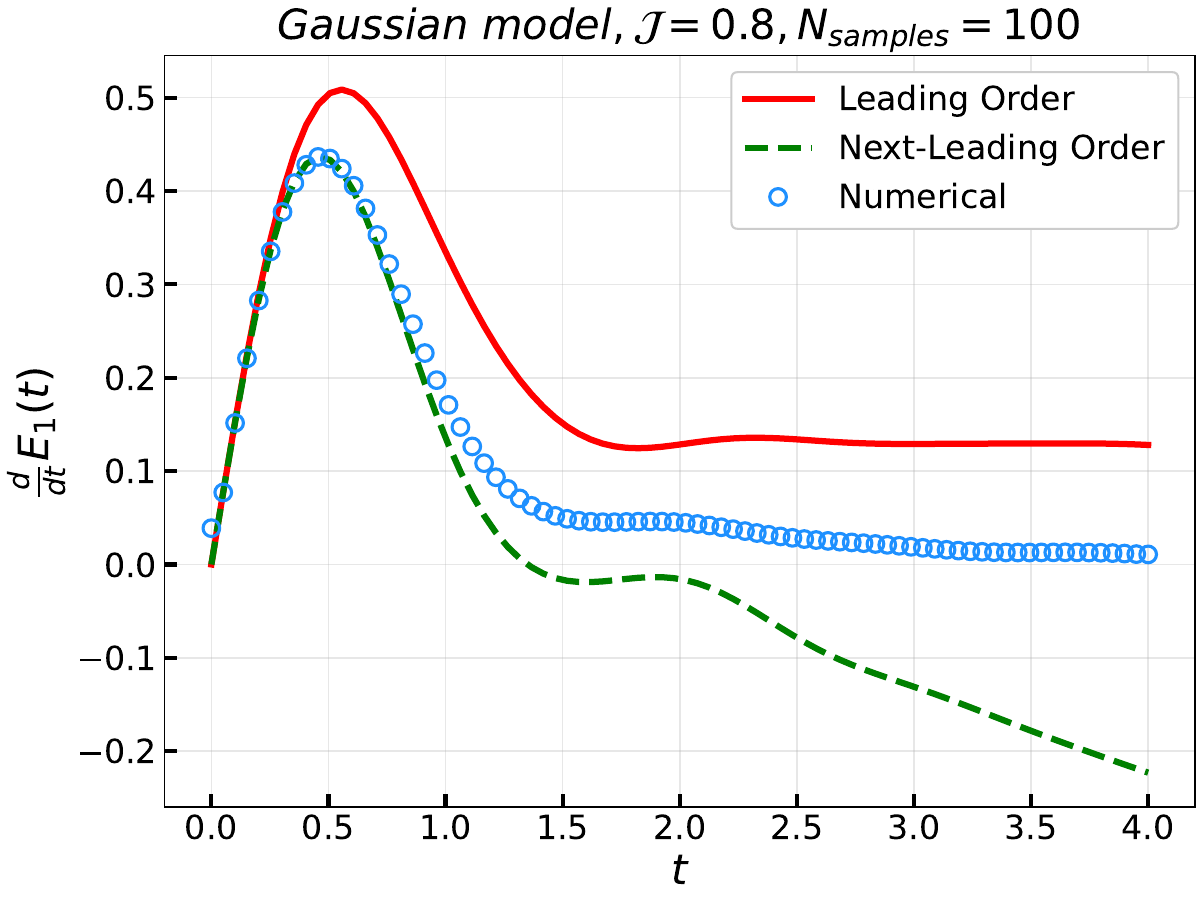}
	\caption{
		Comparison between perturbation theory and numerical simulation for the energy $E_1$ and its time derivative $\dot{E}_1$ of subsystem~1.
		Three coupling strengths are shown: $\mathcal{J}=0.2$, $0.5$, and $0.8$ (from top to bottom).
		The perturbation curves include the leading-order result and the result with second-order correction.
		Numerical data are averaged over 100 realizations of the random coupling.
		Model parameters: $N_1=N_2=10$, $\beta_1=2$, $\beta_2=1$, and both subsystem spectra are drawn from a standard Gaussian distribution ($\mathcal{E}=1$).
		As $\mathcal{J}$ increases, the leading-order prediction deviates more from the numerics, but the second-order correction clearly reduces the error.
		At early times the perturbative results agree well with the simulation for all $\mathcal{J}$.
		For larger couplings, the second-order correction becomes important at later times.
	}
	\label{fig:gaussian_simulation}
\end{center}
\end{figure}
\begin{figure}[htbp]
\begin{center}
	\includegraphics[width=0.48\textwidth]{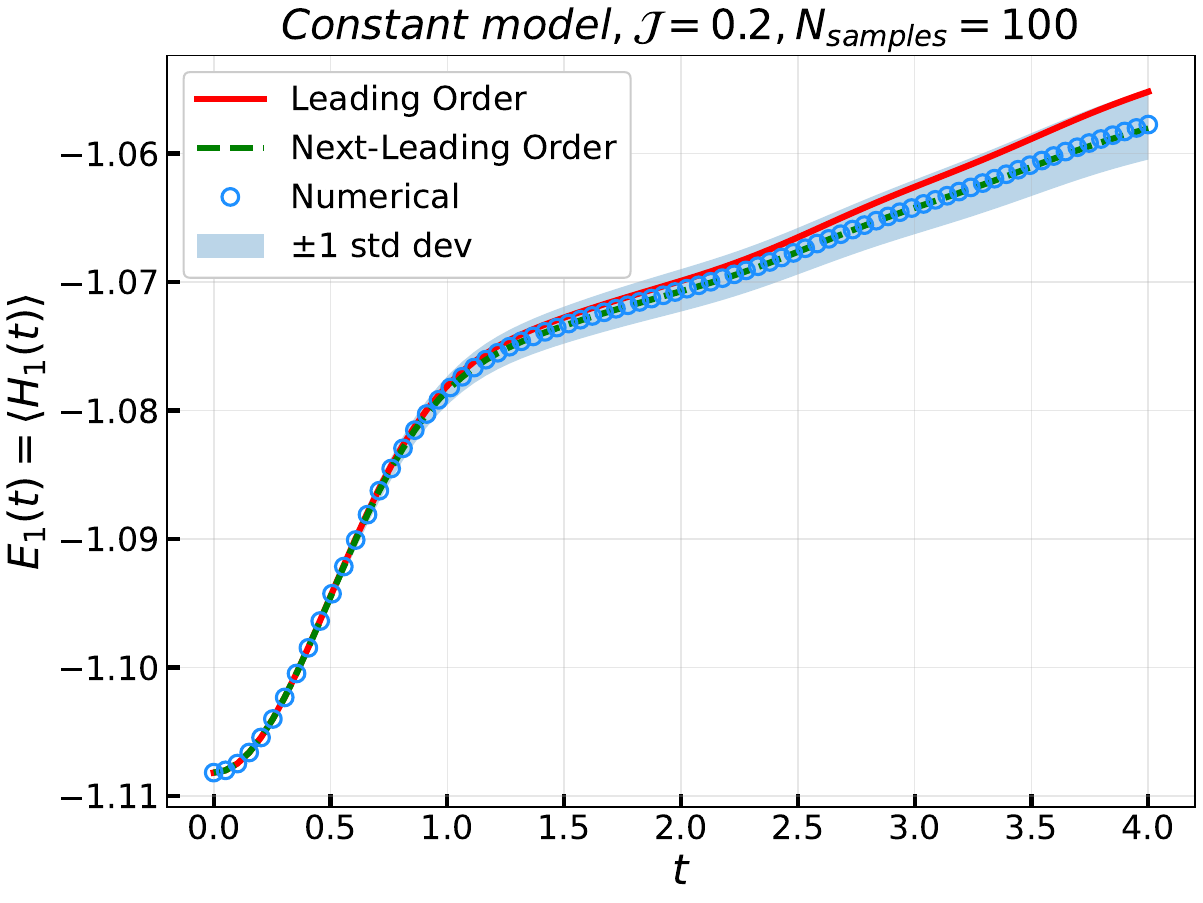}
	\includegraphics[width=0.48\textwidth]{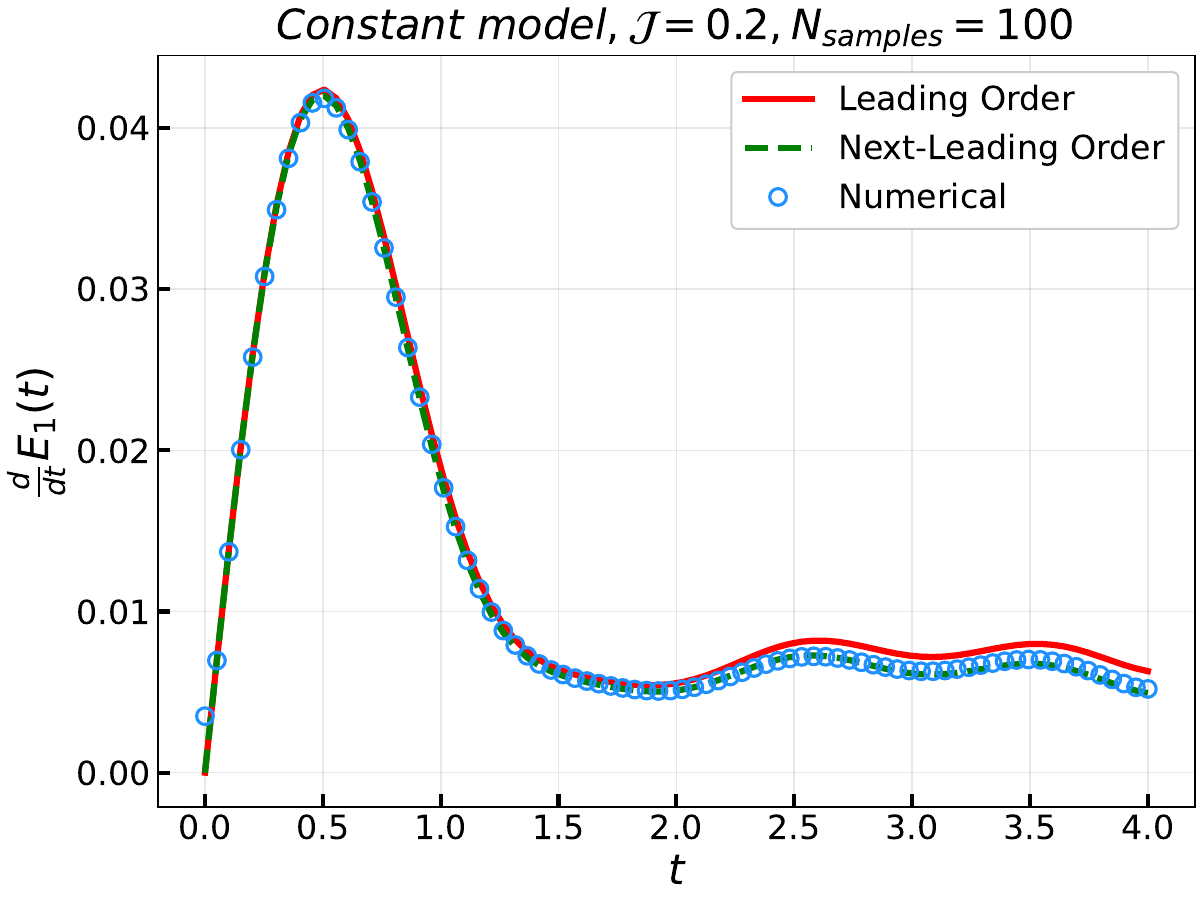}
	\includegraphics[width=0.48\textwidth]{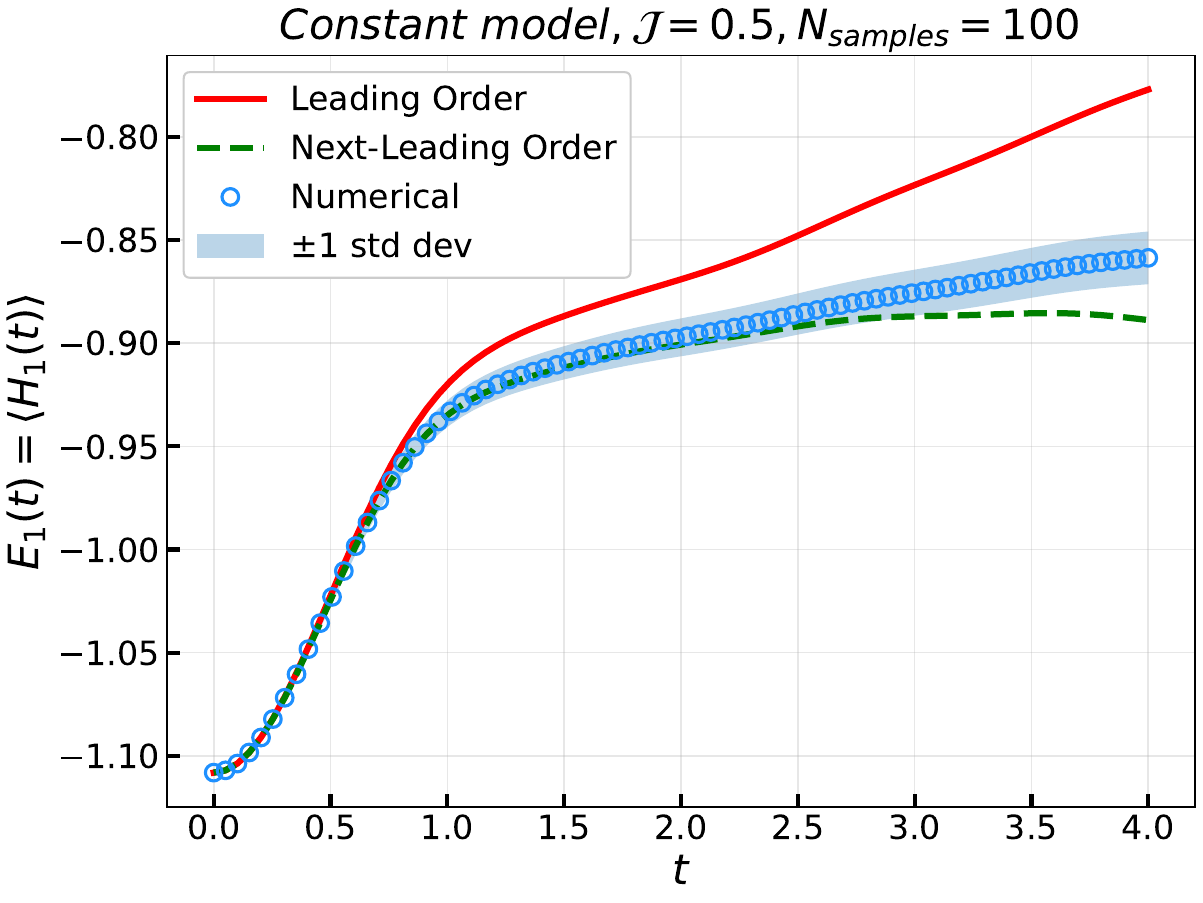}
	\includegraphics[width=0.48\textwidth]{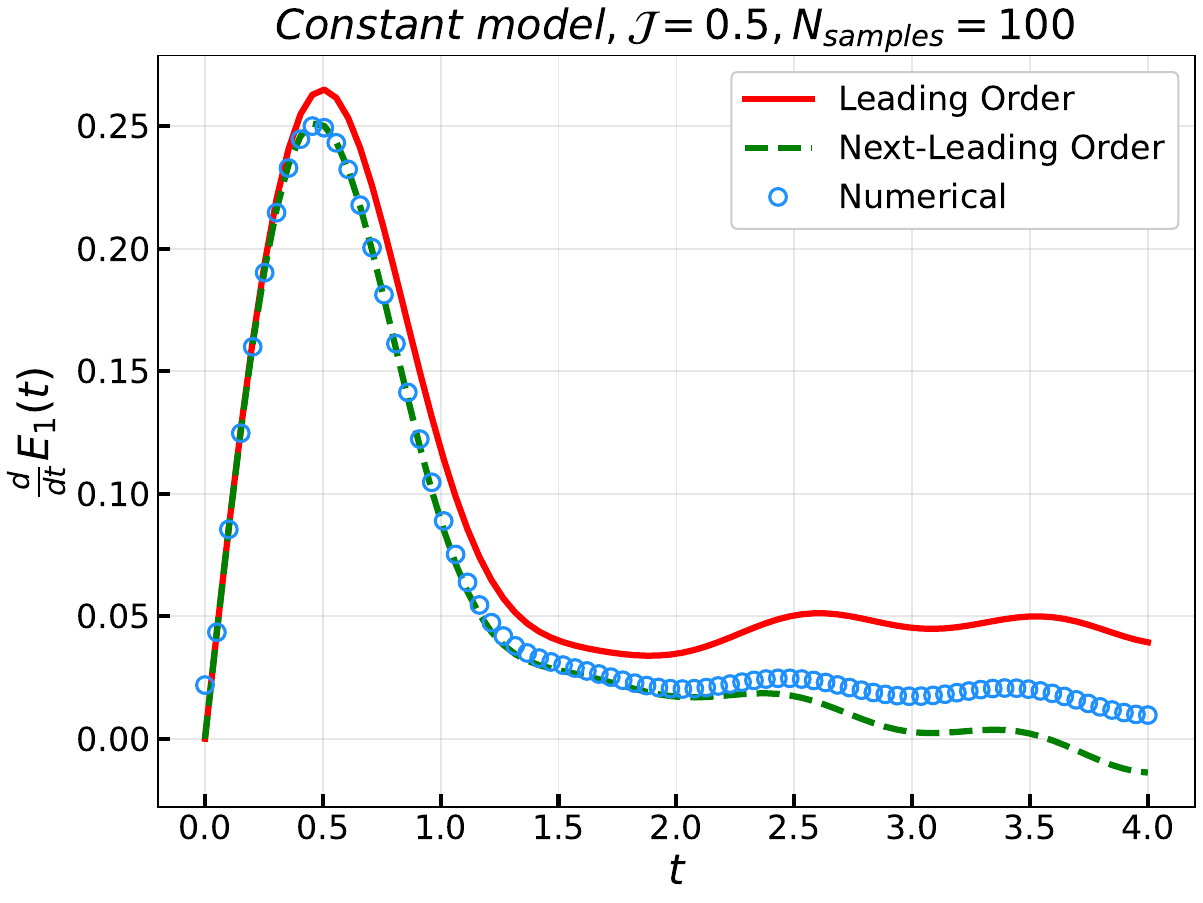}
	\includegraphics[width=0.48\textwidth]{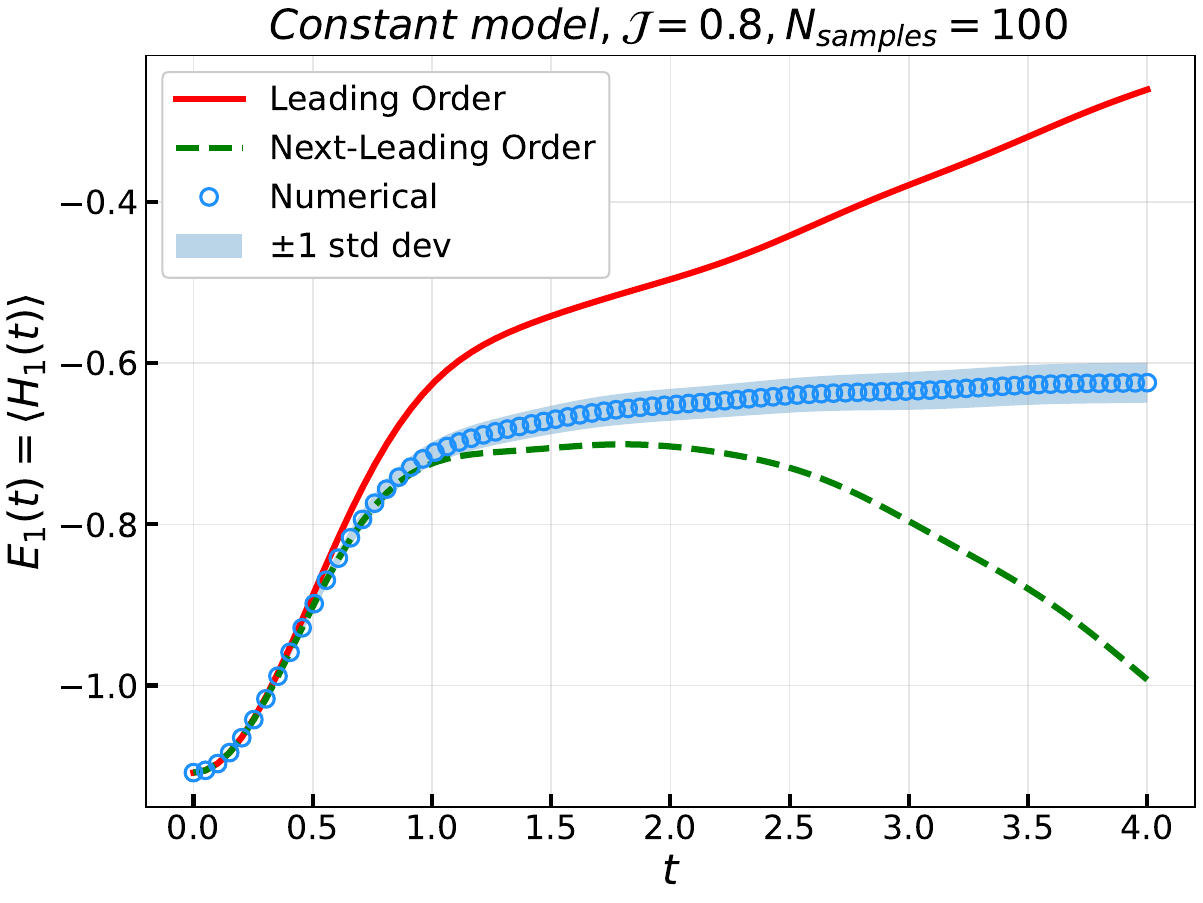}
	\includegraphics[width=0.48\textwidth]{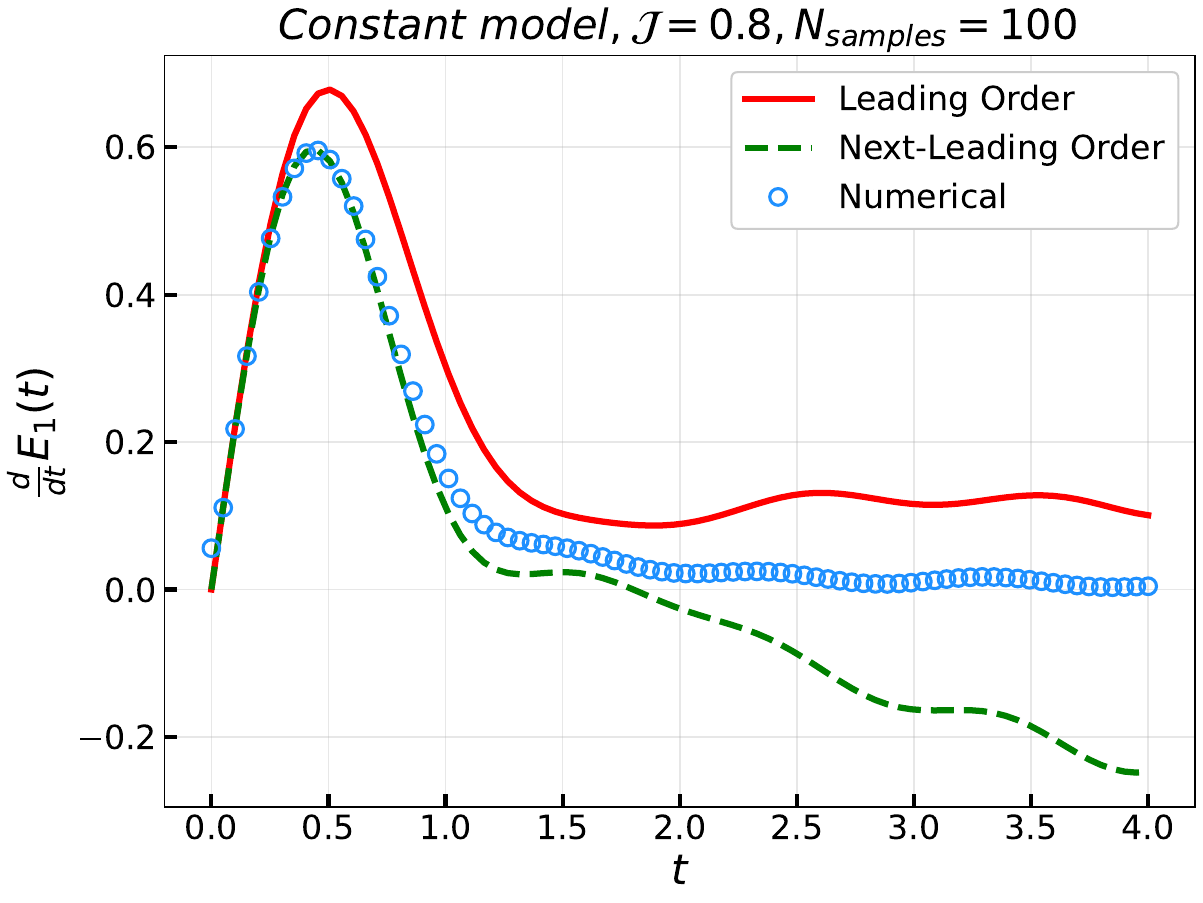}
	\caption{
		Comparison of perturbative predictions with numerical simulations for $E_1$ and $\dot{E}_1$ in the constant-density model.
		Results are shown for $\mathcal{J}=0.2$, $0.5$, and $0.8$ (from top to bottom).
		Model parameters: $N_1=N_2=10$, $\beta_1=2$, $\beta_2=1$, and both subsystem spectra are drawn from a constant distribution (with $\mathcal{E}=1$).
		Numerical data are averaged over 100 realizations of the random coupling.
	}
	\label{fig:const_simulation_const}
\end{center}
\end{figure}
Based on the first- and second-order perturbative results, the general $n$-th-order correction takes the form
\begin{align}
\dot{E}_{1}^{(n)}(t)=\frac{\mathcal{J}^{2n}}{\mathcal{E}^{2n}}f_{n}(\beta_{1},\beta_{2},\mathcal{E},t)\co
\end{align}
where the dimensionless function $f_n$ satisfies
\begin{align}
f_{n}(\beta_{1},\beta_{2},\mathcal{E},t)\big|_{t=0}=0,\qquad f_{n}(\beta_{1},\beta_{2},\mathcal{E},t)\big|_{t=\infty}=\text{finite}\ed
\end{align}
Thus, within the prethermalization plateau the perturbation series is controlled by the small parameter $\mathcal{J}/\mathcal{E}$, and the theory should be accurate when $\mathcal{J}/\mathcal{E}\ll 1$.

To check this, we simulate systems with $N_{1}=N_{2}=10$ and average over $100$ realizations of the random coupling.
In the second-order calculation we simply drop terms with vanishing energy denominators; these contributions are suppressed by $1/N$ in the spectral sum and have negligible effect on the result.
Figs~\ref{fig:gaussian_simulation} and~\ref{fig:const_simulation_const} compare the perturbative predictions with the numerical data for the Gaussian and constant spectral densities.

For the smallest coupling $\mathcal{J}=0.2$, the leading-order result already matches the simulations well at all times shown.
When $\mathcal{J}$ increases to $0.5$ and $0.8$, the leading-order curves begin to deviate, especially at later times $\mathcal{J}t\gtrsim1$, but including the second-order correction clearly reduces the error.
At early times $\mathcal{J}t\ll1$, the perturbative results agree with the numerics for all coupling strengths.
These observations confirm that our perturbative expansion correctly describes the energy dynamics, and that the true expansion parameter is $\mathcal{J}/\mathcal{E}$.

\section{Discussion and Outlook}
\label{sec:discussion}

In this paper, we investigated energy transport between two quantum subsystems coupled by a Gaussian random matrix.
By working in the interaction picture and employing a diagrammatic expansion, we obtained analytic expressions for the energy change rate up to second order in the coupling strength $\mathcal{J}$.
The key physical ingredient of our analysis is the separation of the total energy change into work and heat contributions, based on the decomposition of the total Hamiltonian $\mathcal{H}=H+T$ into commuting and non-commuting parts with respect to $e^{+i\mathcal{H}t}He^{-i\mathcal{H}t}$.
We proved that the heat current defined in this way flows from the hotter subsystem to the colder one at early times, in agreement with the second law of thermodynamics, even though the coupling does not conserve the total energy.
The apparent early-time energy gain of the hotter subsystem is entirely accounted for by the work performed by the non-commuting part of the interaction and does not constitute anomalous heat flow.

For several representative spectral densities, including Gaussian, constant, semicircle, and Gamma distributions, we evaluated the leading-order energy current and the steady-state heat conductance explicitly.
The results exhibit a universal linear growth at short times, a peak, and saturation to a constant value at long times.
The second-order correction was derived in full generality; despite its algebraic complexity, it noticeably improves the agreement with numerical simulations for moderate couplings, confirming that the perturbative series is controlled by the small parameter $\mathcal{J}/\mathcal{E}$.
The numerical data also show that the second-order expression remains accurate up to $\mathcal{J}/\mathcal{E}\lesssim 0.8$.

In the present model the random coupling matrix does not commute with the total Hamiltonian, so the total energy is not strictly conserved.
A complete thermodynamic description of energy transport would require a coupling that preserves the total energy, for instance by projecting onto the commuting part $T_{\parallel}$.
Extending our perturbative framework to such energy-conserving random interactions, or to factorized couplings of the form $T = O_1 O_2$, is a natural direction for future research.
It would also be valuable to investigate finite-$N$ corrections and to connect the present formalism to holographic or SYK-like models, where similar bounds on thermalization have been conjectured.

\section*{Acknowledgments}
We thank Yanyuan Li for helpful discussions regarding this project.
\appendix
\section{Sign of the Heat Current at Early and Late Times}
\label{appdix:sign}

We now verify that the definitions given in Eqs.~\eref{eq:dotQ_smallt} and \eref{eq:dotE1_larget} provide a consistent definition of the heat current by examining their signs.  Specifically, we show that the sign of each expression is the same as \(\beta_1-\beta_2\), so that the heat current obeys the second law of thermodynamics.

Define
\begin{align}
	\Phi(\epsilon_1,\epsilon_2,\eta)=\rho_1(\epsilon_1)\rho_2(\epsilon_2)\rho_1(\epsilon_1+\eta)\rho_2(\epsilon_2-\eta)\ge0\co
\end{align}
and consider the integral without the positive constant:
\begin{align}
	\mathcal{K}= \int d\epsilon_1 d\epsilon_2 d\eta\,\Phi(\epsilon_1,\epsilon_2,\eta)\,
	e^{-\beta_1\epsilon_1-\beta_2\epsilon_2}\,\eta\ed
\end{align}
The sign of the original quantity is exactly that of \(\mathcal{K}\); we therefore only need to show that \(\operatorname{sgn}(\mathcal{K})=\operatorname{sgn}(\beta_1-\beta_2)\).
Make the change of variables
\begin{align}
	\epsilon_1'=\epsilon_1+\eta,\qquad \epsilon_2'=\epsilon_2-\eta,\qquad \eta'=-\eta,
\end{align}
whose Jacobian has absolute value \(1\).  Then \(\epsilon_1=\epsilon_1'+\eta',\;\epsilon_2=\epsilon_2'-\eta'\) and
\begin{align}
	\mathcal{K} &= \int d\epsilon_1' d\epsilon_2' d\eta'\,
	\rho_1(\epsilon_1'+\eta')\rho_2(\epsilon_2'-\eta')\rho_1(\epsilon_1')\rho_2(\epsilon_2')\,
	e^{-\beta_1(\epsilon_1'+\eta')-\beta_2(\epsilon_2'-\eta')}\,(-\eta') \nonumber\\
	&= - \int d\epsilon_1' d\epsilon_2' d\eta'\,
	\Phi(\epsilon_1',\epsilon_2',\eta')\,
	e^{-\beta_1\epsilon_1'-\beta_2\epsilon_2'}\,
	e^{-(\beta_1-\beta_2)\eta'}\,\eta'.
\end{align}
Renaming dummy variables back to \(\epsilon_1,\epsilon_2,\eta\) gives
\begin{align}
	\mathcal{K} = - \int d\epsilon_1 d\epsilon_2 d\eta\,\Phi\,
	e^{-\beta_1\epsilon_1-\beta_2\epsilon_2}\,
	e^{-(\beta_1-\beta_2)\eta}\,\eta.
\end{align}
Adding the original expression for \(\mathcal{K}\) and the one obtained after the change of variables yields
\begin{align}
	2\mathcal{K} &= \int d\epsilon_1 d\epsilon_2 d\eta\,\Phi\,
	e^{-\beta_1\epsilon_1-\beta_2\epsilon_2}\,
	\eta\Bigl[1-e^{-(\beta_1-\beta_2)\eta}\Bigr] \nonumber\\
	&= \int d\epsilon_1 d\epsilon_2 d\eta\,\Phi\,
	e^{-\beta_1\epsilon_1-\beta_2\epsilon_2}\,
	2\eta e^{-\frac{\beta_1-\beta_2}{2}\eta}
	\sinh \left(\frac{\beta_1-\beta_2}{2}\eta\right).
\end{align}
Thus
\begin{align}
	\mathcal{K} = \int d\epsilon_1 d\epsilon_2 d\eta\,\Phi\,
	e^{-\beta_1\epsilon_1-\beta_2\epsilon_2}\,
	e^{-\frac{\beta_1-\beta_2}{2}\eta}\,
	\eta \sinh \left(\frac{\beta_1-\beta_2}{2}\eta\right).
\end{align}
In this final form the integrand factors \(\Phi\ge0\) and the exponentials are strictly positive.
Hence the sign of the whole integral is governed by
\begin{align}
	\eta \sinh \left(\frac{\beta_1-\beta_2}{2}\eta\right).
\end{align}
If \(\beta_1>\beta_2\), then \(\frac{\beta_1-\beta_2}{2}>0\) and \(\eta\sinh(c\eta)>0\) for all \(\eta\neq0\).
If \(\beta_1<\beta_2\), then \(\frac{\beta_1-\beta_2}{2}<0\) and \(\eta\sinh(c\eta)<0\) for all \(\eta\neq0\).
If \(\beta_1=\beta_2\), the factor vanishes identically, giving \(\mathcal{K}=0\).
The integrand does not change sign on the integration domain, except on a set of measure zero.
Therefore
\begin{align}
	\operatorname{sgn}(\mathcal{K})=\operatorname{sgn}(\beta_1-\beta_2),
\end{align}
which completes the proof.
\section{Asymptotic Limits of the Heat Conductance}
\label{appdix:asym}
We now consider the general case in which the two subsystems may have different densities of states, $\rho_1(\epsilon)$ and $\rho_2(\epsilon)$, each with finite spectral bandwidth or sufficiently rapid decay at high energy. The characteristic energy scales are denoted by $\mathcal{E}_1$ and $\mathcal{E}_2$, respectively. The heat conductance can be written as
\begin{align}
	\sigma=\frac{\pi\mathcal{J}^{2}}{\mathcal{Z}}\,\beta^{2}\int d\epsilon_{1} d\epsilon_{2} d\eta\;\rho_{1}(\epsilon_{1})\rho_{2}(\epsilon_{2})\,\rho_{1}(\epsilon_{1}+\eta)\,\rho_{2}(\epsilon_{2}-\eta) e^{-\beta(\epsilon_{1}+\epsilon_{2})}\,\eta\,\bigl(\epsilon_{1}-\epsilon_{2}\bigr)\co\label{eq:sigma-general12}
\end{align}
where $\mathcal{Z}=\mathcal{Z}_{1}\mathcal{Z}_{2}$ with $\mathcal{Z}_{i}=\int_{0}^{\infty}\rho_{i}(\epsilon)e^{-\beta\epsilon}d\epsilon$.

\subsubsection*{High-temperature limit $\beta\mathcal{E}_{1},\beta\mathcal{E}_{2}\ll1$}
In the high-temperature regime the exponential factor is slowly varying, and we expand the integrand for small $\beta$.
To leading order $\mathcal{Z}\to\bigl(\int_{0}^{\infty}\rho_{1}(\epsilon)d\epsilon\bigr)\bigl(\int_{0}^{\infty}\rho_{2}(\epsilon)d\epsilon\bigr)=1$.
Inside the integral we replace $e^{-\beta(\epsilon_{1}+\epsilon_{2})}$ by $1+\mathcal{O}(\beta^{1})$.
One finds
\begin{align}
	\sigma \approx C\,\mathcal{J}^{2}\,\overline{\mathcal{E}}\,\beta^{2}\co
\end{align}
where $\overline{\mathcal{E}}$ is an appropriate average energy scale (e.g.\ $\overline{\mathcal{E}}=(\mathcal{E}_{1}+\mathcal{E}_{2})/2$) and the dimensionless coefficient
\begin{align}
	C=\pi\int d\epsilon_{1}d\epsilon_{2}d\eta\;\frac{\eta\,(\epsilon_{1}-\epsilon_{2})}{\overline{\mathcal{E}}}\,\rho_{1}(\epsilon_{1})\rho_{2}(\epsilon_{2})\,\rho_{1}(\epsilon_{1}+\eta)\,\rho_{2}(\epsilon_{2}-\eta)\,\label{eq:C_general_12}
\end{align}
depends only on the shapes of the spectral densities.
Thus, for arbitrary $\rho_{1},\rho_{2}$ the high-temperature conductance obeys the universal scaling $\sigma\propto\mathcal{J}^{2}\overline{\mathcal{E}}\beta^{2}$.

\subsubsection*{Low-temperature limit $\beta\mathcal{E}_{1},\beta\mathcal{E}_{2}\gg1$}
At low temperatures the integrals are dominated by the neighbourhood of the ground states, which we assume to be located at $\epsilon=0$ for both subsystems (the spectra are bounded from below).
Suppose the densities of states behave as power laws near the edge,
\begin{align}
	\rho_{1}(\epsilon) \simeq A_{1}\,\epsilon^{\nu_{1}}\co\qquad
	\rho_{2}(\epsilon) \simeq A_{2}\,\epsilon^{\nu_{2}}\co\qquad \nu_{1},\nu_{2}\ge0\co
\end{align}
with constants $A_{1},A_{2}$.
Then the partition functions give
\begin{align}
	\mathcal{Z}_{i} \approx A_{i}\int_{0}^{\infty}\epsilon^{\nu_{i}}e^{-\beta\epsilon}d\epsilon
	= A_{i}\,\Gamma(\nu_{i}+1)\,\beta^{-(\nu_{i}+1)}\co
\end{align}
so that $\mathcal{Z}\approx A_{1}A_{2}\Gamma(\nu_{1}+1)\Gamma(\nu_{2}+1)\,\beta^{-(\nu_{1}+\nu_{2}+2)}$.
Introducing rescaled variables $x_{i}=\beta\epsilon_{i}$, $y=\beta\eta$, the numerator of~\eqref{eq:sigma-general12} becomes
\begin{align}
	\begin{aligned}
		&\int d\epsilon_{1}  d\epsilon_{2} d\eta\;\rho_{1}(\epsilon_{1})\rho_{2}(\epsilon_{2})\,\rho_{1}(\epsilon_{1}+\eta)\,\rho_{2}(\epsilon_{2}-\eta)\,e^{-\beta(\epsilon_{1}+\epsilon_{2})}\,\eta\,(\epsilon_{1}-\epsilon_{2})\\\simeq&\;A_{1}^{2}A_{2}^{2}\,\beta^{-2\nu_{1}-2\nu_{2}-5}\int_{0}^{\infty}dx_{1}dx_{2}dy\;x_{1}^{\nu_{1}}x_{2}^{\nu_{2}}(x_{1}+y)^{\nu_{1}}(x_{2}-y)^{\nu_{2}}\,e^{-(x_{1}+x_{2})}\,y\,(x_{1}-x_{2}).
	\end{aligned}
\end{align}
The remaining multiple integral is a pure number that depends only on $\nu_{1},\nu_{2}$.
Assembling the factors yields
\begin{align}
	\sigma \propto \frac{\beta^{2}}{\beta^{-(\nu_{1}+\nu_{2}+2)}}\;
	\beta^{-2\nu_{1}-2\nu_{2}-5}
	= \beta^{-(\nu_{1}+\nu_{2}+1)}.
\end{align}
Restoring the energy scales one may write
\begin{align}
	\sigma\propto\frac{\mathcal{J}^{2}\beta}{\left(\beta\mathcal{E}_{1}\right)^{\nu_{1}+1}\left(\beta\mathcal{E}_{2}\right)^{\nu_{2}+1}}\ed
\end{align}
These results unify the examples analysed in the main text.

\section{Second-order Calculation of the Energy Current}
\label{appdix:NLorder}
This appendix details the derivation of the ensemble-averaged energy of subsystem 1 to second order in the coupling \(\mathcal{J}\), denoted \(\overline{E}_1^{(2)}(t)\) and given in Eq.~\eqref{eq:E1NL}. According to the Dyson series expansion, the second-order contribution consists of terms where the total number of interaction vertices in \(U\) and \(U^\dagger\) sums to four:
\begin{align}
	\overline{E}_1^{(2)}(t)=\sum_{j+k=4}\mathbb{E}\left[\text{Tr}\left(U^{\dagger(j)}(t)H_{1}U^{(k)}(t)\rho_{init}\right)\right]\ed
\end{align}
where the superscripts \((j)\) and \((k)\) denote the perturbative orders of \(U^\dagger\) and \(U\), respectively. The three distinct pairing combinations \((j,k) = \{(0,4)+(4,0), (1,3)+(3,1), (2,2)\}\) are evaluated separately using the Wick contraction rule for the Gaussian random matrix \(T\), \(\mathbb{E}(TAT) = J (\mathrm{Tr} A) \mathbb{I}\), while neglecting the crossing contraction, which is suppressed by the system size.

\paragraph{The (0,4)+(4,0) contribution}
For the \((0,4)\) term, a direct application of the Dyson series and Wick contractions yields
\begin{equation}
	\begin{aligned}
		&\mathbb{E}\left[\text{Tr}\left(U^{\dagger(0)}(t)H_{1}U^{(4)}(t)\rho_{init}\right)\right]\\=&\mathbb{E}\left(e^{i\mathcal{H}_{0}t}H_{1}(-i)^{4}\int_{0}^{t}dt_{>}^{4}\left(\prod_{k=1}^{4}e^{-i\mathcal{H}_{0}(t_{k-1}-t_{k})}\mathcal{H}_{I}\right)e^{-i\mathcal{H}_{0}t_{4}}\right)\\=&J^{2}\int_{0}^{t}d^{4}t_{>}\left[\text{Tr}(A_{1})\text{Tr}(A_{3})\text{Tr}(A_{2}A_{4})+\text{Tr}(A_{2})\text{Tr}(A_{4})\text{Tr}(A_{1}A_{3})\right]\co
	\end{aligned}
\end{equation}
where the operators \(A_i\) are defined as
\begin{equation}
	\begin{aligned}
		A_{1}&=U^{(0)}(t_{4})\rho_{init}U^{\dagger(0)}(t)H_{1}U^{(0)}\left(t-t_{1}\right),~A_{2}=U^{(0)}(t_{1}-t_{2}),\\
		A_{3}&=U^{(0)}(t_{2}-t_{3}),~A_{4}=U^{(0)}(t_{3}-t_{4})\ed
	\end{aligned}
\end{equation}
The contribution from the \((4,0)\) term is obtained by complex conjugation. After performing the time integrals, the combined contribution is 
\begin{align}\label{eq:04}
	\overline{E}_1^{(0,4)+(4,0)}=\frac{\mathcal{J}^{4}}{\mathcal{Z}}\int d[\epsilon]_{\beta_1\beta_2}d[\eta]d[\xi] \epsilon_{1}\left[\frac{t\sin\left(t\left(\epsilon-\eta\right)\right)}{\left(\eta-\xi\right)\left(\epsilon-\eta\right){}^{2}}+\frac{t\sin\left(t\left(\epsilon-\xi\right)\right)}{\left(\xi-\eta\right)\left(\epsilon-\xi\right){}^{2}}-\frac{t^{2}}{\left(\epsilon-\eta\right)\left(\epsilon-\xi\right)}\right]\ed
\end{align}
\paragraph{The (1,3)+(3,1) contribution}
For the \((1,3)\) term, a similar calculation gives
\begin{equation}
	\begin{aligned}
		&\mathbb{E}\left[\text{Tr}\left(U^{\dagger(1)}(t)H_{1}U^{(3)}(t)\rho_{init}\right)\right]\\
		=&-J^2 \int_{0}^{t}d^{4}t_{>}\left[\text{Tr}(A_{1})\text{Tr}(A_{3})\text{Tr}(A_{2}A_{4})+\text{Tr}(A_{2})\text{Tr}(A_{4})\text{Tr}(A_{1}A_{3})\right],
	\end{aligned}
\end{equation}
with the operators
\begin{equation}
	\begin{aligned}
		A_{1}&=\rho_{init}U^{(0)}(t_{4})U^{\dagger(0)}(t-t_{1}),~A_{2}=U^{\dagger(0)}(t_{1})H_{1}U^{(0)}(t-t_{2}),\\
		A_{3}&=U^{(0)}(t_{2}-t_{3}),~A_{4}=U^{(0)}\left(t_{3}-t_{4}\right).
	\end{aligned}
\end{equation}
The \((3,1)\) term follows by conjugation.
\paragraph{The (2,2) contribution}
Finally, for the \((2,2)\) case, we obtain
\begin{equation}
	\begin{aligned}
		&\mathbb{E}\left[\text{Tr}\left(U^{\dagger(2)}(t)H_{1}U^{(2)}(t)\rho_{init}\right)\right]\\
		=&J^2 \int_{0}^{t}d^{4}t_{>}\left[\text{Tr}(A_{1})\text{Tr}(A_{3})\text{Tr}(A_{2}A_{4})+\text{Tr}(A_{2})\text{Tr}(A_{4})\text{Tr}(A_{1}A_{3})\right],
	\end{aligned}
\end{equation}
where
\begin{equation}
	\begin{aligned}
		A_{1}&=\rho_{init}U^{(0)}(t_{4})U^{\dagger(0)}(t-t_{1}),~A_{2}=U^{\dagger(0)}(t_{1}-t_{2}),\\
		A_{3}&=U^{\dagger(0)}(t_{2})H_1U^{(0)}(t-t_{3}),~A_{4}=U^{(0)}\left(t_{3}-t_{4}\right).
	\end{aligned}
\end{equation}
It is convenient to split the \((2,2)\) result into a part symmetric under the exchange of the energy variables \(\xi\) and \(\eta\), denoted \((2,2)_{\text{Sym}}\), and the remaining asymmetric part, \((2,2)_{\text{Asym}}\). The symmetric part evaluates to
\begin{equation}\label{eq:22sym}
	\begin{aligned}
		E_1^{(2,2)_\text{Sym}}=\frac{\mathcal{J}^{4}}{\mathcal{Z}}&\int\frac{d[\epsilon]_{\beta_{1}\beta_{2}}d[\eta]d[\xi]\epsilon_{1}}{(\epsilon-\eta)^{2}(\epsilon-\xi)^{2}}\bigg[\left((\xi-\epsilon)\sin(t(\epsilon-\eta))+(\eta-\epsilon)\sin(t(\epsilon-\xi))\right)t\\&+\left(\epsilon^{2}-\eta\epsilon-\xi\epsilon+\eta\xi\right)t^{2}+1+\cos(t(\eta-\xi))-\cos(t(\epsilon-\eta))-\cos(t(\epsilon-\xi))\bigg]\ed
	\end{aligned}
\end{equation}
\paragraph{Combining all terms}
The asymmetric part of the \((2,2)\) term is combined with the \((1,3)+(3,1)\) contributions for brevity. The result of this combination is
\begin{equation}\label{eq:13p22asym}
	\begin{aligned}
		E_{1}^{(1,3)+(3,1)+(2,2)_{\text{Asym}}}&=\frac{\mathcal{J}^{4}}{\mathcal{Z}}\int d[\epsilon]_{\beta_{1}\beta_{2}}d[\eta]d[\xi]\eta_{1}\bigg[-\frac{2}{(\eta-\xi)(\epsilon-\eta)^{2}(\epsilon-\xi)}\\&-\frac{2t\sin(t(\epsilon-\eta))}{(\eta-\xi)(\epsilon-\eta)(\epsilon-\xi)}-\frac{2\cos(t(\epsilon-\xi))}{(\eta-\xi)^{2}(\epsilon-\eta)(\epsilon-\xi)}+\frac{2\cos(t(\xi-\eta))}{(\eta-\xi)(\epsilon-\eta)(\epsilon-\xi)^{2}}\\&+\frac{2\left(\epsilon^{2}+\eta^{2}+\xi^{2}-\epsilon\eta-\epsilon\xi-\eta\xi\right)\cos(t(\epsilon-\eta))}{(\eta-\xi)^{2}(\epsilon-\eta)^{2}(\epsilon-\xi)^{2}}\bigg]\ed
	\end{aligned}
\end{equation}
Summing the expressions in Eqs.~\eqref{eq:04}, \eqref{eq:22sym}, and \eqref{eq:13p22asym} yields the complete second-order result, \(\overline{E}_1^{(2)}(t)\), presented in Eq.~\eqref{eq:E1NL} of the main text.

\section{Evaluation of the Steady-State Integral for the Constant Distribution}
\label{appdix:derive}
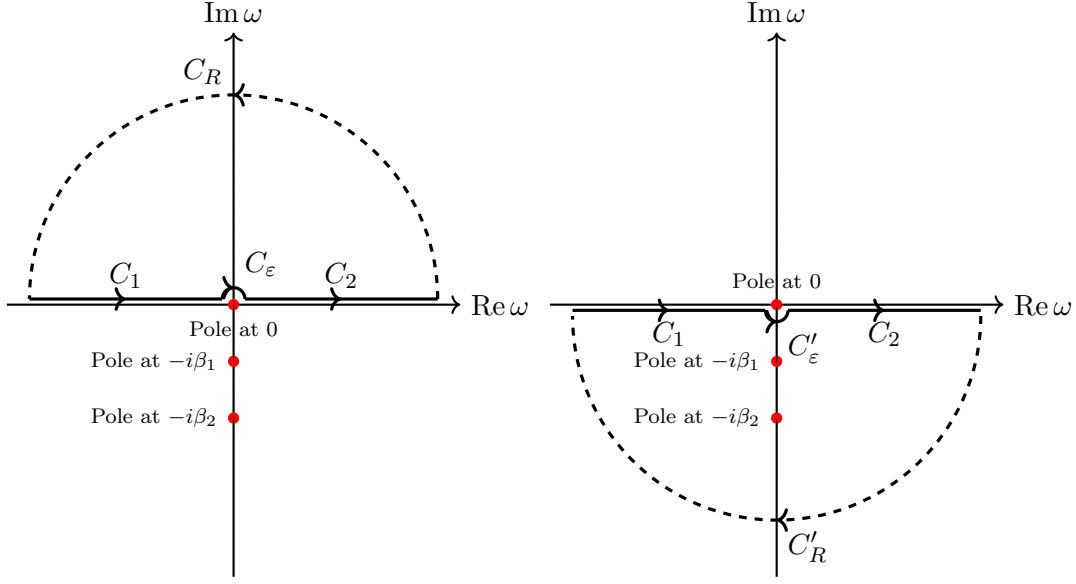
\begin{figure}[ht]
\begin{center}
	\begin{tikzpicture}[scale=1.5,
		pole/.style={circle, fill=red, inner sep=1.5pt},
		arrow/.style={decoration={markings, mark=at position 0.5 with {\arrow{>}}}, postaction={decorate}}
		]
		
		\draw[->, thick] (-2, 0) -- (2, 0) node[right] {$\mathrm{Re}\,\omega$};
		\draw[->, thick] (0, -2.4) -- (0, 2.4) node[above] {$\mathrm{Im}\,\omega$};
		
		\node[pole, label=below:{\scriptsize Pole at $0$}] at (0,0) {};
		\node[pole, label=left:{\scriptsize Pole at $-i\beta_1$}] at (0,-0.5) {};
		\node[pole, label=left:{\scriptsize Pole at $-i\beta_2$}] at (0,-1) {};
		
		\draw[arrow, very thick] (-1.8, 0.05) -- (-0.1, 0.05) node[midway, above] {$C_1$};
		\draw[arrow, very thick] (0.1, 0.05) -- (1.8, 0.05) node[midway, above] {$C_2$};
		
		\draw[arrow, very thick] (-0.1, 0.05) arc (180:0:0.1) node[midway, above right] {$C_\varepsilon$};
		
		\draw[arrow, very thick, dashed] (1.8, 0.05) arc (0:180:1.8) node[midway, above left] {$C_R$};
		
		
		\draw[dashed] (0,-1.2) -- (0,1.5);
		
	\end{tikzpicture}
	\begin{tikzpicture}[scale=1.5,
		pole/.style={circle, fill=red, inner sep=1.5pt},
		arrow/.style={decoration={markings, mark=at position 0.5 with {\arrow{>}}}, postaction={decorate}}
		]
		
		\draw[->, thick] (-2, 0) -- (2, 0) node[right] {$\mathrm{Re}\,\omega$};
		\draw[->, thick] (0, -2.4) -- (0, 2.4) node[above] {$\mathrm{Im}\,\omega$};
		
		\node[pole, label=above:{\scriptsize Pole at $0$}] at (0,0) {};
		\node[pole, label=left:{\scriptsize Pole at $-i\beta_1$}] at (0,-0.5) {};
		\node[pole, label=left:{\scriptsize Pole at $-i\beta_2$}] at (0,-1) {};
		
		\draw[arrow, very thick] (-1.8, -0.05) -- (-0.1, -0.05) node[midway, below] {$C_1$};
		\draw[arrow, very thick] (0.1, -0.05) -- (1.8, -0.05) node[midway, below] {$C_2$};
		
		\draw[arrow, very thick] (-0.1, -0.05) arc (180:360:0.1) node[midway, below right] {$C_\varepsilon'$};
		
		\draw[arrow, very thick, dashed] (1.8, -0.1) arc (0:-180:1.8) node[midway, below right] {$C_R'$};
		
		
		\draw[dashed] (0,-1.5) -- (0,1.2);
		
	\end{tikzpicture}
	\caption{Integration contours in the complex \(\omega\)-plane for \(I_+\) (left) and \(I_-\) (right). The red dots indicate the poles at \(\omega = 0, -i\beta_1, -i\beta_2\).}
	\label{fig:contours}
\end{center}
\end{figure}

This appendix provides the detailed derivation of the steady-state energy current for the constant distribution, as presented in Eq.~\eqref{eq:current_const}. Starting from the frequency representation of the current, we obtain 
\begin{align}
	\dot{\overline{E}}^{(1)}_{1}(\infty)&=\frac{\beta_{1}\beta_{2}\mathcal{J}^{2}}{32\mathcal{E}^{2}\sinh\left(2\beta_{1}\mathcal{E}\right)\sinh\left(2\beta_{2}\mathcal{E}\right)}\int_{-\infty}^{+\infty}\frac{1-\cos(4\mathcal{E}\omega)}{\omega^{2}\left(\omega+i\beta_{1}\right){}^{2}\left(\omega+i\beta_{2}\right){}^{2}}\nn
	&\newline\times \bigg[4\mathcal{E}\left(\omega+i\beta_{1}\right)\left(\omega+i\beta_{2}\right)\sinh\left(2\left(\beta_{1}-\beta_{2}\right)\mathcal{E}\right)+\left(\beta_{1}-\beta_{2}\right)\cosh\left(2\mathcal{E}\left(\beta_{1}+\beta_{2}-2i\omega\right)\right)\nn
	&\newline\newline+\left(\beta_{2}-\beta_{1}\right)\cosh\left(2\left(\beta_{1}-\beta_{2}\right)\mathcal{E}\right)\bigg]\nn
	&\equiv \frac{\beta_{1}\beta_{2}\mathcal{J}^{2}}{32\mathcal{E}^{2}\sinh\left(2\beta_{1}\mathcal{E}\right)\sinh\left(2\beta_{2}\mathcal{E}\right)} \mathcal{I}\ed
\end{align}
The integral to be evaluated is
\begin{align}
		\mathcal{I}=\int_{-\infty}^{+\infty}d\omega(I_++I_-)
\end{align}
where the integrands \(I_+\) and \(I_-\) are given by
\begin{align}
I_+&=\frac{e^{-2\left(\beta_{1}+\beta_{2}\right)\mathcal{E}}}{4\omega^{2}\left(\beta_{1}-i\omega\right){}^{2}\left(\omega+i\beta_{2}\right){}^{2}}
\bigg[8\mathcal{E}(z-2)\left(\omega+i\beta_{1}\right)\left(\omega+i\beta_{2}\right)e^{2\left(\beta_{1}+\beta_{2}\right)\mathcal{E}}\sinh\left(2\left(\beta_{1}-\beta_{2}\right)\mathcal{E}\right)\nn
&\newline+\left(\beta_{1}-\beta_{2}\right)\left(e^{4\left(\beta_{1}+\beta_{2}\right)\mathcal{E}}+(z-1)^{2}\right)-2\left(\beta_{1}-\beta_{2}\right)(z-2)e^{2\left(\beta_{1}+\beta_{2}\right)\mathcal{E}}\cosh\left(2\left(\beta_{1}-\beta_{2}\right)\mathcal{E}\right)\bigg]\co
\end{align}
\begin{align}
I_-&=\frac{1}{4\omega^{2}z^{2}\left(\omega+i\beta_{1}\right){}^{2}\left(\omega+i\beta_{2}\right){}^{2}}\bigg[-8\mathcal{E}z\left(\omega+i\beta_{1}\right)\left(\omega+i\beta_{2}\right)\sinh\left(2\left(\beta_{1}-\beta_{2}\right)\mathcal{E}\right)\nn
&\newline+\left(\beta_{1}-\beta_{2}\right)(2z-1)e^{2\left(\beta_{1}+\beta_{2}\right)\mathcal{E}}+2\left(\beta_{1}-\beta_{2}\right)z\cosh\left(2\left(\beta_{1}-\beta_{2}\right)\mathcal{E}\right)\bigg]\ed
\end{align}
Here we have introduced the shorthand \(z = e^{4 i \mathcal{E} \omega}\).
The integral \(\mathcal{I}\) is evaluated via contour integration in the complex \(\omega\)-plane. As illustrated in Fig.~\ref{fig:contours}, different contours are chosen for \(I_+\) and \(I_-\) to ensure convergence. For \(I_+\), the contour is closed in the upper half-plane, while for \(I_-\), it is closed in the lower half-plane. 
By Cauchy's theorem, the integrals over the closed contours are
\begin{align}
0&=\oint I_{+}d\omega=\int_{C_{1}+C_{2}}I_{+}d\omega+\int_{C_{\epsilon}}I_{+}d\omega+\int_{C_{R}}I_{+}d\omega\co\nn
-2\pi i\sum_{\omega=-i\beta_{1},-i\beta_{2}}\text{Res}\left(I_{-}\right)&=\oint I_{-}d\omega=\int_{C_{1}+C_{2}}I_{-}d\omega+\int_{C_{\epsilon}'}I_{-}d\omega+\int_{C_{R}'}I_{-}d\omega\ed
\end{align}
In the limit of an infinite contour radius, the contributions from the large arcs \(C_R\) and \(C_R'\) vanish. The desired integral \(\mathcal{I}\) is then given by the sum of the integrals along the real axis, \(\int_{C_1+C_2} (I_+ + I_-) d\omega\). Combining the two Cauchy relations, we obtain
\begin{align}
\mathcal{I}=-2\pi i\sum_{\omega=-i\beta_{1},-i\beta_{2}}\text{Res}\left(I_{-}\right)-\int_{C_{\epsilon}}I_{+}d\omega-\int_{C_{\epsilon}'}I_{-}d\omega\ed
\end{align}
The final result, Eq.~\eqref{eq:current_const}, is obtained by evaluating the residues at the poles \(\omega = -i\beta_1\) and \(\omega = -i\beta_2\) and the two small semicircular integrals around the pole at \(\omega = 0\).

\bibliographystyle{JHEP}
\bibliography{ref}
\end{document}